\documentclass[a4paper,11pt]{article}
\usepackage{jcappub} % for details on the use of the package, please
\usepackage{amsmath,amssymb}
\usepackage{color,graphicx}
\usepackage{setspace}
\date{}

\def\bea{\begin{eqnarray}}
\def\eea{\end{eqnarray}}
\def\be{\begin{equation}}
\def\ee{\end{equation}}
\def\ba{\begin{eqnarray}}
\def\ea{\end{eqnarray}}
\def\beq{\begin{eqnarray}}
\def\eeq{\end{eqnarray}}

\def\d{\mathrm{d}}

\newcommand{\Hm}{{\mathcal{H}}}
\newcommand{\HH}{{\mathcal{H}}}

\newcommand{\DE}{{\rm DE}}
\newcommand{\DM}{{\rm DM}}
\newcommand{\G}{\zeta}
\newcommand{\diff}{{\rm d}}

\title{On structure formation from a small-scales-interacting dark sector} 
	\author[1]{Mahnaz Asghari,}
	\author[2]{Jose Beltr\'an Jim\'enez,}
    \author[1]{Shahram Khosravi,}
	\author[3]{David F. Mota}
		
    \affiliation[1]{Department of Astronomy and High Energy Physics, Faculty of Physics,\\ Kharazmi University, Tehran, Iran.}
	\affiliation[2]{Departamento de F\'isica Fundamental and IUFFyM, Universidad de Salamanca,\\ E-37008 Salamanca, Spain.}
	\affiliation[3]{Institute of Theoretical Astrophysics, University of Oslo, 0315 Oslo, Norway.}	
	
\emailAdd{std\_m.asghari@khu.ac.ir}	
	\emailAdd {jose.beltran@usal.es}
	\emailAdd{khosravish@khu.ac.ir}
	\emailAdd{mota@astro.uio.no}

\abstract{We consider a cosmological model with an interaction between dark matter and dark energy which leaves the background cosmology unaffected and only affects the evolution of the perturbations. This is achieved by introducing a coupling given in terms of the relative velocities of dark matter and dark energy. This interaction has the distinctive feature of appearing predominantly on small scales, where peculiar velocities can become important. We confront the predictions of the model to cosmological observations and find a potential alleviation of the known tension in the amplitude of density perturbations as measured by low redshift galaxy surveys and the Planck data. The model also predicts a shift in the turnover of the matter power spectrum which does not depend on the horizon at equality (fixed by the background cosmology and, thus, unaffected by the perturbations) and is entirely due to the interaction between dark matter and dark energy. A bias in the peculiar velocity between baryons and dark matter is also shown to be a unique feature of this type of interactions in the dark sector.}


\begin{document}
\maketitle
\flushbottom
\section{Introduction}
\label{sec:intro}

 In the last decades, cosmological measurements from Type Ia supernovae (SNeIa) \cite{1538-3881-116-3-1009,0004-637X-517-2-565}, the Cosmic Microwave Background (CMB) anisotropies \cite{2014A&A...571A...1P,2014A&A...571A..16P,2016A&A...594A..13P}, Large Scale Structures (LSS) \cite{PhysRevD.69.103501,1538-3881-128-1-502,1538-3881-129-3-1755}, Integrated Sachs-Wolfe (ISW) effect \cite{1967ApJ...147...73S,1538-4357-683-2-L99,2004Natur.427...45B}, and weak lensing \cite{PhysRevLett.90.221303}, strongly indicate that the universe is presently undergoing an accelerated expansion and established $\Lambda$CDM as the standard model of cosmology. This model depicts a present universe largely dominated by Dark Energy (DE), in the form of a cosmological constant, and Dark Matter (DM), required to be sufficiently cold as to drive the formation of structures today. This model thus conforms a universe dominated by a dark sector whose true nature however remains elusive, but its success lies in its ability to explain most of the above cosmological observations. It has however some theoretical problems (such as the cosmological constant problem \cite{RevModPhys.61.1,Martin:2012bt} or the coincidence problem \cite{Zlatev:1998yg,ArkaniHamed:2000tc,Dodelson:2001fq,Malquarti:2003hn,Jimenez:2008er,Jimenez:2008au,Velten2014}) that have triggered an intense activity in the development of models for both DM and DE. Moreover, from the observational point of view, there are some unexplained tensions arising from several observational probes: One is the tension between the measured $\sigma_8$ from low redshift galaxy surveys \cite{doi:10.1046/j.1365-8711.2003.06550.x} which is lower than the inferred $\sigma_8$ obtained from the CMB dataset from Planck \cite{2016A&A...594A..13P}.  The second important tension arises from local measurements of the Hubble constant \cite{0004-637X-826-1-56,0004-637X-775-1-13} which are in conflict with the predictions made from the Planck CMB observations \cite{2016A&A...594A..13P}. These discrepancies can be due to systematic effects in the methods used for measurements \cite{doi:10.1093/mnras/stw707,PhysRevLett.110.241305}, or they could indeed indicate that our cosmological model does not perfectly describe the observed universe and could signal physics beyond $\Lambda$CDM. Considering the latter possibility, in this paper we explore a model for the dark sector which may alleviate the differences between low redshift surveys and high redshift observations. Our model is based on a dark matter component which is coupled to dark energy via a non-gravitational interaction and this coupling can indeed help alleviating the $\sigma_8$ tension, while it will have nothing to say about the tension in the Hubble constant because the background evolution of the dark sector remains unaltered.

Models for interacting dark energy - dark matter have been proposed for the first time long ago. The main original motivation was to alleviate the coincidence problem because, for a coupled system, it is more natural to have attractor solutions where all components have comparable density parameters. On the other hand, our uncertainty on the underlying mechanisms for both DM and DE leads to the reasonable doubt to what extent these components are completely decoupled (other than gravitationally) and, thus, to the logical exploration of interactions in the dark sector. There are many different models according to the type of coupling between DM and DE, the nature of the DE field, the choice of the self-interacting potential for the DE, etc. \cite{PhysRevD.79.123506,1475-7516-2010-05-009,PhysRevD.84.023010, PhysRevD.91.063530,PhysRevD.77.063513,Cruz2008338,PhysRevD.79.063518,MohseniSadjadi2010,1475-7516-2009-07-027,BALDI2009178,1475-7516-2012-04-011,1475-7516-2013-01-010,PhysRevD.87.044018}. A large number of studies were conducted to study the features and signatures of interacting models both at cosmological and astrophysical scales  \cite{PhysRevD.89.103531,Brevik2014,PhysRevD.92.123537,1475-7516-2015-07-036,1475-7516-2015-07-015,PhysRevD.92.043524,Brevik2015,mota1,mota2,1475-7516-2016-04-014,PhysRevD.94.043518,PhysRevD.94.103526,doi:10.1093/mnras/stw1087,1475-7516-2017-01-028,1475-7516-2016-12-009,Nozari201692,PhysRevD.94.123511,doi:10.1093/mnras/stw2073,Neomenko2016,PhysRevD.95.043513,1475-7516-2017-01-050,PhysRevD.95.043526,Ebrahimi2016,PhysRevD.95.023515,doi:10.1093/mnras/stw3358} (see also \cite{Wang:2016lxa} for a review). 

While the large majority of the coupled dark energy models have a strong effect at both the background cosmology and at the perturbation level, the model we explore here has the remarkable feature that the interaction in the dark sector only affects the inhomogeneous perturbations, while the background cosmology remains unmodified. An interaction similar to the one we will explore here was considered in \cite{Simpson:2010vh} (see also \cite{Baldi:2016zom,Kumar:2017bpv}), where a possible elastic scattering between DM and DE was suggested. There it was already recognised that interactions in the dark sector can leave the background cosmology unaffected. The invoked mechanism was a DM-DE interaction giving rise to an elastic scattering with momentum transfer but without energy transfer, similar to the Thomson scattering of photons and baryons before recombination. We will find similar properties for our interaction in this work. Another framework where it was noticed the possibility of having a dark sector that only interacts at the perturbative level was the general scenario for interacting DM and DE developed in \cite{Pourtsidou:2013nha}, where they dubbed this type of interaction pure momentum exchange. See also \cite{Skordis:2015yra} for a general parameterized Post-Friedmann (PPF) framework of an interacting dark sector where DE is described by a scalar field. 

The approach that we will follow in this work is more phenomenological than the ones mentioned in the previous paragraph, but it can be readily related to them. One advantage of our framework is that it will allow for more directly pinpointing the effects entirely due to the interaction, for instance by keeping the sound speed of DE equal to one. Of course, this comes at the expense of having less theoretically well-motivated models. In any case, our different approach will complement the ones already studied in the literature and we will seize the opportunity to comment on some interesting consequences that have not been sufficiently stressed thus far. Our phenomenological model-building framework will be based on directly modifying the conservation equations of the dark sector, while Einstein equations and ordinary matter conservation equations will remain the same. More explicitly, we will construct our interaction by introducing a coupling given in terms of the relative velocities of DM and DE. As explained above, this interaction has the remarkable feature that will only appear on small scales, where peculiar velocities can become important, but the super-horizon (as well as the background) cosmology remains the same. This is so thanks to the cosmological principle according to which the large scale frame of all the components should coincide, i.e., peculiar velocities should tend to zero as we take larger and larger scales. The possibility that dark energy could have a different large scale rest frame from the rest of components has been explored in e.g. \cite{Maroto:2005kc,BeltranJimenez:2007rsj,Jimenez:2008vs,Harko:2013wsa} (see also the recent paper \cite{Cembranos:2019plq} with a very thorough analysis of cosmologies with non-comoving fluids). In these models, our interaction would have effects also at the background level. However, since we are interested in maintaining the background unaffected we will not consider this possibility here, although it could also give rise to an interesting and distinctive phenomenology. The nice feature of our class of interactions is that it allows to modify the evolution of the cosmological perturbations from purely sub-horizon physics (interactions in the dark sector) and not at the expense of changing some cosmological parameters. This is somewhat similar to the Effective Field Theory of DE \cite{Gubitosi:2012hu} where the background and the perturbations are conveniently disentangled. A similar approach was followed in \cite{Barros:2018efl} where an interacting quintessence model conformally coupled to matter was considered and where the background is fixed to be the same as in $\Lambda$CDM. Notice however the difference with the class of interacting models discussed above and that we will consider in this work, where the background is not affected by the interaction by construction and without the need of fixing it to exhibit any particular behaviour (e.g. $\Lambda$CDM).  As a consequence, these interactions open up the possibility of modifying the growth of structures for a given amount of matter fixed by the background cosmology, which is oblivious to the interaction in the dark sector. The feasibility of this mechanism was in fact shown within the elastic scattering of DM-DE model  \cite{Baldi:2016zom} and the PPF frameworks \cite{PhysRevD.94.043518}. We will confirm this tendency with the model under consideration in this work.

This paper is organized as follows: In section \ref{sec2} we introduce the interacting dark energy model. We study initial conditions and evolution of dark matter perturbations analytically in section \ref{sec3}. In section \ref{sec4} we discuss the most interesting features obtained from the numerical solutions. Section \ref{sec4.4} is devoted to confronting the model to cosmological observations. Finally, we discuss our conclusions in section \ref{sec5}.

\section{Interacting model} \label{sec2}

In this section we will introduce the model that we will use throughout the present work. In a general model featuring interactions between DM and DE, the usual conservation equations are modified to
\be \label{eq1}
\nabla_\mu T^{\mu\nu}_{(\DM)}=Q^\nu,\quad\quad\nabla_\mu T^{\mu\nu}_{(\DE)}=-Q^\nu \,,
\ee
where $Q^\nu$ describes the interaction between both components and we impose conservation of the DM-DE system so that $\nabla_\mu(T^{\mu\nu}_{(\DM)}+T^{\mu\nu}_{(\DE)})=0$. This means that the equations for photons, baryons and neutrinos will remain the same. Furthermore, we will assume that both DM and DE components are well-described by a perfect fluid so that their energy-momentum tensors will be given by
\be \label{eq5}
T_{(i)}^{\mu\nu}=\big(\rho_{(i)}+p_{(i)}\big)\,u_{(i)}^\mu\,u_{(i)}^\nu+g^{\mu\nu}\,p_{(i)}\,,\quad\quad i={\rm DM, DE} 
\ee
with $\rho_{(i)}$ and $p_{(i)}$ their corresponding energy densities and pressures and $u_{(i)}$ their comoving 4-velocities. Since we want an interaction that only affects the inhomogeneous perturbations while the background cosmology remains unmodified, a natural choice is a coupling given in terms  of the relative velocities of dark matter $u^\mu_{(\DM)}$ and dark energy $u^\mu_{(\DE)}$ so we will assume
\be \label{eq2}
Q^\mu=\alpha \,\Big(u^\mu_{(\DE)}-u^\mu_{(\DM)}\Big) \,,
\ee
with $\alpha$ measuring the strength of the interaction. In principle, $\alpha$ can be a function of time (e.g. it can be a function of the densities and/or pressures of the fluids), but for the sake of simplicity we will assume it to be a constant parameter (at least for the scales of interest to us).  Since the large scale rest frames of dark matter and dark energy are assumed to be the same, this interaction will only appear on small scales, where peculiar velocities can become important. The background cosmology will be assumed to be that of a spatially flat Friedmann-Lema\^itre-Robertson-Walker (FLRW) universe described by the line element
\be \label{eq3}
\d s^2=-\d t^2+a^2(t)\d\vec{x}^2=a^2(\tau)(-\d\tau^2+\d \vec{x}^2)\,,
\ee
expressed in cosmic $t$ and conformal $\tau$ time respectively. We then have the usual conservation equations for the background 
\be \label{eq4}
\bar{\rho}'_{(i)}+3\Hm(\bar{\rho}_{(i)}+\bar{p}_{(i)})=0\,,\quad\quad i=\DM,\;\DE
\ee
which give $\bar{\rho}_{(\DM)}\propto a^{-3}$ and $\bar{\rho}_{(\DE)}\propto e^{-3\int(1+w_{\mathrm{(DE)}})\diff N}$ with $w_{\mathrm{(DE)}}=\bar{p}_{(\DE)}/\bar{\rho}_{(\DE)}$ the dark energy equation of state parameter, $N=\log a$ the number or e-folds variable, and $\mathcal{H}=a'/a$ the conformal Hubble parameter (a bar stands for a quantity evaluated on the background and a prime indicates derivative with respect to conformal time). As advertised, this evolution is oblivious to the interaction and only the perturbations will be sensitive to it. 

On the other hand, the gravitational field equations are not (directly) affected by the interaction between dark matter and dark energy, since we impose conservation of the DM-DE system, so we have the usual Einstein equations
	\begin{equation}  \label{eq6}
	    G_{\mu \nu}=8\pi G\: T_{\mu \nu}  \,,
	\end{equation} 
with $G_{\mu \nu}$ the Einstein tensor and $T_{\mu \nu}$ the total momentum-energy tensor. The background cosmological evolution for the universe containing radiation (R), baryons (B), dark matter and dark energy is thus given by the usual Friedmann equation 
	\begin{equation} \label{eq7}
        \mathcal{H}^2=\frac{8 \pi G}{3}\,a^2\,\Big(\bar{\rho}_\mathrm{(R)}+\bar{\rho}_\mathrm{(B)}+\bar{\rho}_\mathrm{(DM)}+\bar{\rho}_\mathrm{(DE)}\Big)  \,.
	\end{equation}
    
Let us now turn to the more interesting sector of the scalar perturbations (we will not consider vector nor tensor perturbations in this work). We will follow here the notation and conventions of Ref. \cite{Ma:1995ey}. The perturbed metric in the Newtonian gauge is given by
\bea
\d s^2=a^2(\tau)\, \Big[-(1+2\Psi)\d \tau^2+(1-2\Phi)\d \vec{x}^2\Big] \,, \label{eq8}
\eea
with $\Psi$ and $\Phi$ the two gravitational potentials. From the off-diagonal Einstein equations we obtain that these two potentials are equal, which is of course a consequence of working within GR and neglecting any anisotropic stress. It is also useful to consider the synchronous gauge, which is defined by the perturbed line element being
\bea
\d s^2=a^2(\tau)\,\Big[-\d\tau^2+(\delta_{ij}+h_{ij})\d x^i\d x^j\Big] \,,\label{eq9}
\eea
where the perturbed spatial metric is written in terms of the scalar perturbations $h$ and $\eta$ as $h_{ij}={\rm diag}(-2\eta,-2\eta,h+4\eta)$, where, without loss of generality, we assume that the perturbations propagate in the $z$-direction (or, in other words, the corresponding wave vector points along the $z$-direction). 

The perturbations of the dark energy fluid will be characterized by its sound speed $c_s^2$ that relates the pressure and the density perturbations. The proper definition of this quantity, for it to be gauge invariant, is in the rest frame (rf) of the fluid so that (see e.g. \cite{Kodama:1985bj,Hu:1998kj,Bean:2003fb})
\be \label{eq10}
c_{s\mathrm{(DE)}}^2\equiv\left(\frac{\delta p_{(\DE)}}{\delta \rho_{(\DE)}}\right)_{\rm rf} \,,
\ee
where the rest frame of the fluid is defined as the one where $T^0{}_i=0$. In an arbitrary frame we then have
\be \label{eq11}
\delta p_{\mathrm{(DE)}}=c_{s\mathrm{(DE)}}^2\,\delta \rho_{\mathrm{(DE)}}+3\,\Hm\, (1+w_{\mathrm{(DE)}})\,(c_{s\mathrm{(DE)}}^2-c_{a\mathrm{(DE)}}^2)\,\bar{\rho}_{\mathrm{(DE)}}\,\frac{\theta_{\mathrm{(DE)}}}{k^2} \,,
\ee
with
\be \label{eq12}
c_{a\mathrm{(DE)}}^2\equiv\frac{\bar{p}'_{\mathrm{(DE)}}}{\bar{\rho}'_{\mathrm{(DE)}}}= w_{\mathrm{(DE)}}-\frac{w'_{\mathrm{(DE)}}}{3\Hm(1+w_{\mathrm{(DE)}})} \,,
\ee
the adiabatic sound speed. Notice that the second equality also holds for our interacting model because the interaction only appears at the perturbative level, while the background evolution of the fluids is the usual one described by Eqs (\ref{eq4}). In  general, if the interaction has a non-vanishing background contribution $\bar{Q}^0$, the above expression should be modified to
\be \label{eq12+}
c_{a\mathrm{(DE)}}^2\equiv\frac{\bar{p}'_{\mathrm{(DE)}}}{\bar{\rho}'_{\mathrm{(DE)}}}= w_{\mathrm{(DE)}}-\frac{w'_{\mathrm{(DE)}}}{3\Hm(1+w_{\mathrm{(DE)}})-\bar{Q}_0/\bar{\rho}_\DE} \,.
\ee
Thus, another remarkable property of the interaction used in this work is that also the adiabatic sound speed is oblivious to the interaction, as a consequence of having a vanishing interaction at the background level.

Since the interaction term is determined by the relative velocity of the dark matter and dark energy components, it is useful to give here the expressions for the perturbed 4-velocities, which are given by
\bea
u^\mu&=&\frac{1}{a}\,\big((1-\Psi),\vec{v}_\mathrm{(con)}\big) \,, \label{eq13} \\ 
u^\mu&=&\frac{1}{a}\,\big(1,\vec{v}_\mathrm{(syn)}\big) \,, \label{eq14}
\eea
in conformal (con) and synchronous (syn) gauge respectively. In both cases, we have defined the 3-velocities as $\vec{v}={\rm d} \vec{x}/{\rm d}\tau$ in the corresponding coordinates. As usual, for the scalar part of the velocity perturbation, we will use the variable $\theta\equiv i\vec{k}\cdot\vec{v}$, defined in Fourier space. The important point that we want to stress here is that, under a gauge transformation $\delta x^\mu=\xi^\mu $, the interaction changes with the Lie derivative as $\delta Q^\mu=-\mathcal{\xi} \bar{Q}^\mu$ which is identically zero by virtue of the vanishing $\bar{Q}^\mu$. This implies that the interaction term is gauge invariant to first order and, therefore, it is a physical quantity. This will be useful when interpreting the numerical results. 

Due to the gauge invariance of the interaction at first order, we can compute it in any gauge and this will be valid in all gauges. For instance, in the synchronous gauge we have that the temporal component of the 4-velocities of the fluids are not perturbed and, consequently, we will have that $\delta Q^0=0$. This means that the zero components of the first order conservation equations will remain the same and only the momentum equations will be affected. For this reason, this type of interaction was dubbed pure momentum exchange in \cite{PhysRevD.94.043518} . Thus, the conservation equations in synchronous gauge are 
\begin{align}
\delta'_{(\DM)\mathrm{syn}}=&-\left(\theta_{(\DM)\mathrm{syn}}+\frac12 h'\right) \,,  \label{eq15} \\
\theta'_{(\DM)\mathrm{syn}}=&-\HH\theta_{(\DM)\mathrm{syn}}+\zeta\Big(\theta_{(\DE)\mathrm{syn}}-\theta_{(\DM)\mathrm{syn}}\Big) \,, \label{eq16}\\
\delta'_{(\DE)\mathrm{syn}}=&-(1+w_{\mathrm{(DE)}})\Big(\theta_{(\DE)\mathrm{syn}}+\frac12 h'\Big)-3\big(c^2_{s\mathrm{(DE)}}-w_{\mathrm{(DE)}}\big)\HH\,\delta_{(\DE)\mathrm{syn}} \nonumber \\
&-9(1+w_{\mathrm{(DE)}})\big(c^2_{s\mathrm{(DE)}}-w_{\mathrm{(DE)}}\big)\frac{\HH^2}{k^2}\,\theta_{(\DE)\mathrm{syn}} \,, \label{eq17} \\
\theta'_{(\DE)\mathrm{syn}}=&\big(-1+3\,c^2_{s\mathrm{(DE)}}\big)\HH\,\theta_{(\DE)\mathrm{syn}}+\frac{c^2_{s\mathrm{(DE)}}\,k^2}{1+w_{\mathrm{(DE)}}}\,\delta_{(\DE)\mathrm{syn}}-R\zeta\Big(\theta_{(\DE)\mathrm{syn}}-\theta_{(\DM)\mathrm{syn}}\Big) \,,\label{eq18}
\end{align}
while in Newtonian gauge they read
\begin{align}
\delta'_{\mathrm{(DM)con}}=&-\theta_{\mathrm{(DM)con}}+3\,\Phi'  \label{eq19} \,,\\
\theta'_{\mathrm{(DM)con}}=&-\mathcal{H}\,\theta_{\mathrm{(DM)con}}+k^2\,\Phi+\G\Big(\theta_{\mathrm{(DE)con}}-\theta_{\mathrm{(DM)con}}\Big) \label{eq20} \,,\\
\delta'_{\mathrm{(DE)con}}=&-3\,\mathcal{H}\,(c^2_{s\mathrm{(DE)}}-w_{\mathrm{(DE)}})\,\delta_{\mathrm{(DE)con}}-\theta_{\mathrm{(DE)con}}\,(1+w_{\mathrm{(DE)}})\,(1+9\,\mathcal{H}^2\,(c^2_{s\mathrm{(DE)}}-w_{\mathrm{(DE)}})\,\frac{1}{k^2}) \nonumber \\
&+3\,(1+w_{\mathrm{(DE)}})\,\Phi' \label{eq21} \,,\\
\theta'_{\mathrm{(DE)con}}=&(-1+3\,c^2_{s\mathrm{(DE)}})\,\mathcal{H}\,\theta_{\mathrm{(DE)con}}+k^2\,\Phi 
+\frac{k^2\,c^2_{s\mathrm{(DE)}}}{1+w_{\mathrm{(DE)}}}\,\delta_{\mathrm{(DE)con}}-\G R\Big(\theta_{\mathrm{(DE)con}}-\theta_{\mathrm{(DM)con}}\Big) \label{eq22} \,,
\end{align}
where in the above equations and from now on we assume a constant dark energy equation of state parameter, i.e. $w'_{\mathrm{(DE)}}=0$, and we have introduced the quantities
\be
\G\equiv \frac{\alpha\, a}{\bar{\rho}_{(\DM)}}\,,\quad\quad R=\frac{\bar{\rho}_{(\DM)}}{(1+w_{\mathrm{(DE)}})\,\bar{\rho}_{(\DE)}} \,, \label{eq23}
\ee
so that $\G$ and $R\G$ measure the strength of the interaction for dark matter and dark energy, respectively. Let us notice the formal resemblance of these equations with the drag terms proportional to $\zeta$ and those of the photon-baryon plasma before recombination where baryons and photons interact through Thomson scattering. This similarity was taken further in \cite{Simpson:2010vh} where the interaction was assumed to arise through an elastic scattering of DM and DE. Our approach will reduce to the model considered there by assuming a time-dependent interaction with $R\zeta\simeq a n_{\DM} \sigma_{\rm D}$, i.e., the interaction is proportional to the DM abundance $n_{\DM}$ and the corresponding cross-section of the dark sector $\sigma_{\DM}$. Our assumption of a constant coupling parameter $\alpha$ will then lead to a differently evolving $\zeta$. In particular, while the elastic scattering model gives rise to a decaying coupling, our assumption leads to a growing one so that our interaction in fact becomes more important at late times, while at early times it is negligible. However, let us notice that it will still have an effect on the velocities at early times, as we will explain below.

\begin{figure}[htb!]
\centering
\includegraphics[width=14cm]{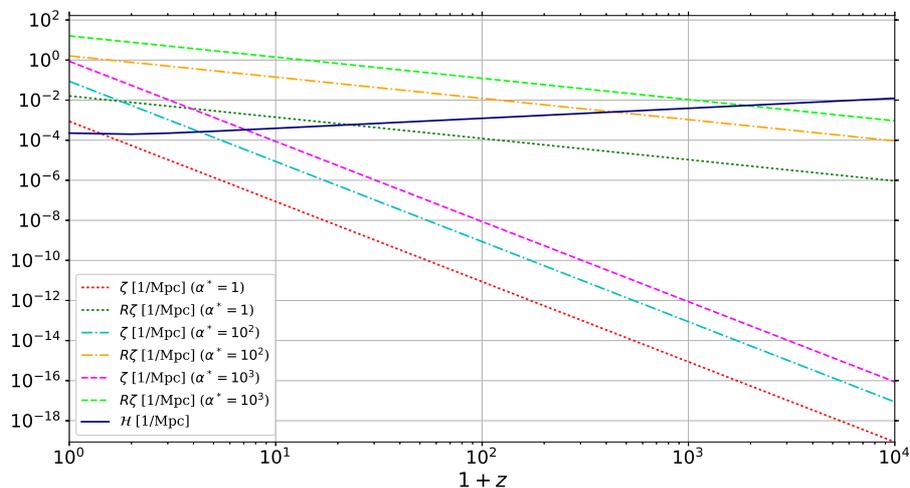}
\caption{In this Figure we show the cosmological evolution of the interaction functions as defined in (\ref{eq23}) for $\Omega_{(\DM),0}=0.26$, $\Omega_{(\rm B),0}=0.04$, $\Omega_{(\DE),0}=0.7$ and three different values of $\alpha^*$. In this figure we can clearly see that the interaction becomes relevant for the dark energy sector much earlier than in the dark matter sector.}
\label{Fig:interaction}
\end{figure}

As discussed above, we see that the interaction only affects the momentum equation and it is the same in both gauges as it should. The effect of the interaction is determined by the parameters introduced in (\ref{eq23}) as compared to the Hubble expansion rate $\Hm$. This ratio can be conveniently expressed as
\be
\frac{\G}{\Hm}=\frac{\alpha^*}{E(z)\,\Omega_{(\DM),0}\,(1+z)^3} \,, \label{eq24}
\ee
where we have normalized the Hubble function to its value today $H_0$ as $E(z)=H(z)/H_0$ and we have introduced the dimensionless parameter
\be \label{eq25}
\alpha^*=\frac{8\pi G\,\alpha}{3\,H_ 0^3}\,.
\ee
We will denote $\tau_\zeta$ the time at which the interaction becomes efficient. This is nothing but the usual competition between the interaction rate and the expansion rate that determines the efficiency of an interaction in an expanding universe. On the other hand, the parameter $R$ measures how much more ($R>1$) or less $(R<1)$ important the interaction is in the dark energy sector. It can be readily expressed in terms of the redshift as
\be \label{eq26}
R=\frac{\Omega_{(\DM),0}}{(1+w_{\mathrm{(DE)}})\,\Omega_{(\DE),0}}\,(1+z)^{-3\,w_{\mathrm{(DE)}}}\,.
\ee
This expression goes as $R\sim (1+w_{\mathrm{(DE)}})^{-1}(1+z)^3\gg 1$ so we see that the interaction will be more important in the dark energy sector than in the dark matter sector throughout the evolution of the universe (see Fig. \ref{Fig:interaction}).

The gravitational Einstein equations will be the same as in the standard non-interacting case, because they do not contain the interaction explicitly. Thus, the linearized Einstein equations in synchronous gauge are
\begin{align}
& k^2\,\eta-\frac{1}{2}\,\HH\,h'=-4\pi G\,a^2\,\sum_{\mathrm{A}} \delta \rho_{\mathrm{(A)syn}} \,, \label{e1} \\
& k^2\,\eta'=4\pi G\,a^2\,\sum_{\mathrm{A}} (\bar{\rho}_{\mathrm{(A)}}+\bar{p}_{\mathrm{(A)}})\,\theta_{\mathrm{(A)syn}} \,, \label{e2} \\
& h''+2\,\HH\,h'-2\,k^2\,\eta=-24\pi G\,a^2\,\sum_{\mathrm{A}} \delta p_{\mathrm{(A)syn}} \,, \label{e3} \\
& h''+6\,\eta''+2\,\HH\,(h'+6\,\eta')-2\,k^2\,\eta=0 \,, \label{e4}
\end{align}
while in the Newtonian gauge they take the form
\begin{align}
& k^2\,\Phi+3\,\HH\,\Phi'+3\,\HH^2\,\Phi=-4\pi G\,a^2\,\sum_{\mathrm{A}} \delta \rho_{\mathrm{(A)con}} \,, \label{e5} \\
& k^2\,\Phi'+\HH\,k^2\,\Phi=4\pi G\,a^2\,\sum_{\mathrm{A}} (\bar{\rho}_{\mathrm{(A)}}+\bar{p}_{\mathrm{(A)}})\,\theta_{\mathrm{(A)con}} \,, \label{e6} \\
& \Phi''+3\,\HH\,\Phi'+(2\HH'+\HH^2)\,\Phi=4\pi G\,a^2\,\sum_{\mathrm{A}} \delta p_{\mathrm{(A)con}} \,, \label{e7}
\end{align}
where the sums over $A$ run over all the components in the universe, i.e., photons, neutrinos, baryons, dark matter and dark energy.

\section{Evolution of the perturbations}	 \label{sec3}

Now that we have introduced the model, let us discuss some of its main features before analysing the numerical solutions. The first thing to keep in mind is that, given the form of our interaction, effects will only appear for sub-Hubble modes, meaning that the super-Hubble modes will evolve as usual in the $\Lambda$CDM model (barring the small modifications owed to considering an equation of state parameter slightly different from $-1$). Thus, we can focus our discussion on the sub-Hubble modes where the effects will be more prominent, except when discussing the initial conditions. As explained above, the interaction becomes relevant for the dark energy much earlier than for the dark matter. However, the density contrast of dark energy is so small that the leading effect actually comes from the interaction in the dark matter sector. We will confirm this later from the numerical solutions. 

\subsection{Behaviour in the radiation epoch: Initial conditions} \label{sec3.1}
We will start our analysis by studying the initial conditions so we will consider the behaviour for modes well outside the horizon and deep in the radiation era. At sufficiently early times, the interaction will be such that $R\zeta\ll \HH$ so we can safely neglect it for now. The only possible issue would be the presence of some growing mode due to the interaction, but that is not the case for the interaction at hand. As usual, the metric perturbation in the synchronous gauge will evolve according to the growing mode
\be \label{eq27}
h=C(k\tau)^2\,,
\ee
where $C$ is the $k-$dependent constant to be matched to the primordial spectrum generated during inflation. The dark matter and dark energy fluids will quickly reach the attractor solution driven by $h$. Since the dark matter velocity perturbations will be very small, we have that the dominant contributions in the dark matter equations are
\bea
\delta_{(\DM)\mathrm{syn}}&\simeq& -\frac12 h=-\frac{C}{2}(k\tau)^2 \,,       \label{eq28} \\
\theta'_{(\DM)\mathrm{syn}}&\simeq& \zeta \theta_{(\DE)\mathrm{syn}}\,.\label{eq29}
\eea
We find the usual growth of the dark matter density contrast and the novel effect due to the interaction that the dark matter velocity is driven by $\zeta\theta_{(\DE)\mathrm{syn}}$. In fact, this is the reason why we cannot use the residual gauge freedom of the synchronous gauge to set $\theta_{(\DM)\mathrm{syn}}=0$ as in the standard non-interacting case. On the other hand, the dark energy perturbations equations in this regime are well approximated by
\bea
\delta'_{(\DE)\mathrm{syn}}&=&-\frac{1+w_{\mathrm{(DE)}}}{2}\, h'-3\,\HH\,(c^2_{s\mathrm{(DE)}}-w_{\mathrm{(DE)}})\delta_{(\DE)\mathrm{syn}}\nonumber\\
&&-9\,(1+w_{\mathrm{(DE)}})\,(c^2_{s\mathrm{(DE)}}-w_{\mathrm{(DE)}})\,\frac{\HH^2}{k^2}\,\theta_{(\DE)\mathrm{syn}}, \\
\theta'_{(\DE)\mathrm{syn}}&=&-(1-3\,c^2_{s\mathrm{(DE)}})\,\HH\,\theta_{(\DE)\mathrm{syn}}+\frac{c^2_{s\mathrm{(DE)}}\,k^2}{1+w_{\mathrm{(DE)}}}\delta_{(\DE)\mathrm{syn}}\,. \label{eq31}
\eea
It is not difficult to see that the attractor of these equations is the  inhomogeneous solution driven by $h$, which reads
\begin{align}
\delta_{(\DE)\mathrm{syn}}\simeq&-(1+w_{\mathrm{(DE)}})\,\frac{4-3\,c^2_{s\mathrm{(DE)}}}{4+3 c^2_{s\mathrm{(DE)}}-6w_{\mathrm{(DE)}}}\,\frac{h}{2}\simeq -\frac{C}{2}\,(1+w_{\mathrm{(DE)}})\frac{4-3c^2_{s\mathrm{(DE)}}}{4+3 c^2_{s\mathrm{(DE)}}-6w_{\mathrm{(DE)}}}\,(k\tau)^2 \label{eq32}\,,\\
\theta_{(\DE)\mathrm{syn}}\simeq &-3\frac{c^2_{s\mathrm{(DE)}}k^2}{4+3 c^2_{s\mathrm{(DE)}}-6w_{\mathrm{(DE)}}}\int \frac{h}{2}\d\tau \simeq -\frac{C}{2}\frac{c^2_{s\mathrm{(DE)}}k}{4+3\, c^2_{s\mathrm{(DE)}}-6w_{\mathrm{(DE)}}}(k\tau)^3 \,,\label{eq33}
\end{align}
This attractor solution was also obtained in \cite{Ballesteros:2010ks}. The solution for the DE velocity perturbations will then source the DM velocity perturbation according to (\ref{eq29}). During radiation we have $\zeta=\alpha^* H_0 \tau^4/\Omega_{(\DM),0}$ so that the dark matter velocity perturbation is given by
\be \label{eq34}
\theta_{(\DM)\mathrm{syn}}\simeq\int \zeta\theta_{(\DE)\mathrm{syn}}\d\tau\simeq  -\frac{C\alpha^*H_0}{16\Omega_{(\DM),0}}\frac{c^2_{s\mathrm{(DE)}}}{4+3 c^2_{s\mathrm{(DE)}}-6w_{\mathrm{(DE)}}}k^4\tau^8 \,,
\ee
where we have neglected the sub-dominant constant mode. We then see how the interaction, even if it is completely negligible deep in the radiation dominated epoch, has the non-trivial effect of sourcing the DM velocity with a very rapidly growing mode $\theta_{(\DM)\mathrm{syn}}\propto \tau^8$. Notice that the dark matter velocity perturbation is greatly suppressed with respect to the dark energy velocity because the interaction $\zeta$ is assumed to be much smaller than $\HH$ in the radiation era. More specifically, from the evolution equation for $\theta_{(\DM)\mathrm{syn}}$ we have
\be \label{eq35}
\theta_{(\DM)\mathrm{syn}}\sim \frac{\zeta}{\HH}\theta_{(\DE)\mathrm{syn}}\ll\theta_{(\DE)\mathrm{syn}}\,,
\ee
deep in the radiation era which in turn confirms the validity of our assumption that $\theta_{(\DM)\mathrm{syn}}\ll\theta_{(\DE)\mathrm{syn}}$ at very early times (see Fig. \ref{velocities}). 

\subsection{Dark matter density contrast}  \label{eq3.2}
We will now focus on the evolution of the density contrast of dark matter, which is in fact the relevant one to study the expected effects in the matter power spectrum. In order to gain some intuition, let us consider the second order differential equation for $\delta_{(\DM)\mathrm{con}}$ in the Newtonian gauge obtained by eliminating the velocity perturbation from Eqs (\ref{eq19}) and (\ref{eq20}) so we have
\be \label{eq36}
\delta_{(\DM)\mathrm{con}}''+\big(\Hm+\zeta\big)\delta_{(\DM)\mathrm{con}}'=3\big(\Hm\Phi'+\Phi''\big)-k^2\Phi+\zeta\big(3\Phi'-\theta_{(\DE)\mathrm{con}}\big)\,.
\ee
The last term in this expression is of the order of the dark energy density contrast, as can be seen in (\ref{eq21}). Since this quantity is negligible throughout the universe evolution, we can safely drop it and finally obtain the result that the dark matter density contrast will evolve according to
\be \label{eq37}
\delta_{(\DM)\mathrm{con}}''+\big(\Hm+\zeta\big)\delta_{(\DM)\mathrm{con}}'\simeq3\big(\Hm\Phi'+\Phi''\big)-k^2\Phi\,,
\ee
which coincides with the usual equation except that now the friction term is given by $\Hm+\zeta$. This means that, as long as $\zeta/\Hm\ll1$, the evolution will mimic that of the standard $\Lambda$CDM cosmology. As we explained above, it is this quantity rather than $\zeta R$ that determines when the interaction will give rise to new effects despite having $R\gg1$. 

By combining Eqs (\ref{eq19}) and (\ref{eq21}), it is possible to find the following equation for sub-Hubble modes
\be \label{eq38}
\delta'_{(\DM)\mathrm{con}}-\frac{\delta'_{(\DE)\mathrm{con}}+3\Hm(c^2_{s\,\mathrm{(DE)}}-w_{\mathrm{(DE)}})\delta_{(\DE)\mathrm{con}}}{1+w}=\theta_{(\DE)\mathrm{con}}-\theta_{(\DM)\mathrm{con}}\,.
\ee
As already mentioned and as we will confirm from our numerical solutions, the density contrast of dark energy perturbations are negligible as compared to the dark matter density contrast so we can neglect the second term of the LHS in the above equation and obtain
\be \label{eq39}
\delta'_{(\DM)\mathrm{con}}\simeq\theta_{(\DE)\mathrm{con}}-\theta_{(\DM)\mathrm{con}}\,.
\ee
This relation means that, once the interaction becomes fully efficient, the two components, namely dark matter and dark energy, are tightly coupled so that they have the same velocity perturbation and, consequently, the density contrast in the dark matter will freeze. This means that, once the interaction locks the DM-DE system, the growth of structures ceases, thus giving rise to a suppression in the matter power spectrum. We will confirm this result below from the numerical solutions.

In the next section we will modify the publicly available Boltzmann code Cosmic Linear Anisotropy Solving System (CLASS) \cite{1475-7516-2011-07-034} to incorporate the effects of the interaction under consideration in this work and we will furthermore use the M\textsc{onte} P\textsc{ython} code \cite{Audren:2012wb,Brinckmann:2018cvx} to obtain the corresponding constraints on the interaction strength.

\section{Numerical results} \label{sec4}
In this section, we will present the results obtained from the modified CLASS code including the interaction. We will use the usual cosmological parameters supplemented by the parameter $\alpha^*$ that measures the strength of the interaction. For the dark energy equation of state parameter we will choose $w_{\mathrm{(DE)}}=-0.98$ in order to fix it and to avoid some divergences in the perturbation equations. Moreover, we will fix its sound speed $c^2_{s\mathrm{(DE)}}=1$. Since we want to study the effects of the interaction, we will keep all the remaining cosmological parameters fixed to the best fit values of $\Lambda$CDM according to Planck 2015 data \cite{2016A&A...594A..13P} given by $\Omega_{(\rm B),0} h^2 = 0.02230$, $\Omega_{(\DM),0} h^2 = 0.1188$, $H_0 = 67.74 $ km/s/Mpc, $A_s = 2.142\times10^{-9}$ and $\tau_{\rm reio} = 0.066$. Let us recall again that the background cosmology is insensitive of the interaction so that all the effects discussed in this section are genuinely due to the coupling in the dark sector.

%%%%%%%%%%%%%%%%%%%%%%%%%%%%%%%%%%
%%%%%%%%%%%%%%%%%%%%%%%%%%%%%%%%%%
\subsection{CMB and matter power spectra} \label{sec4.1}

\begin{figure}[ht!]
	\includegraphics[width=8cm]{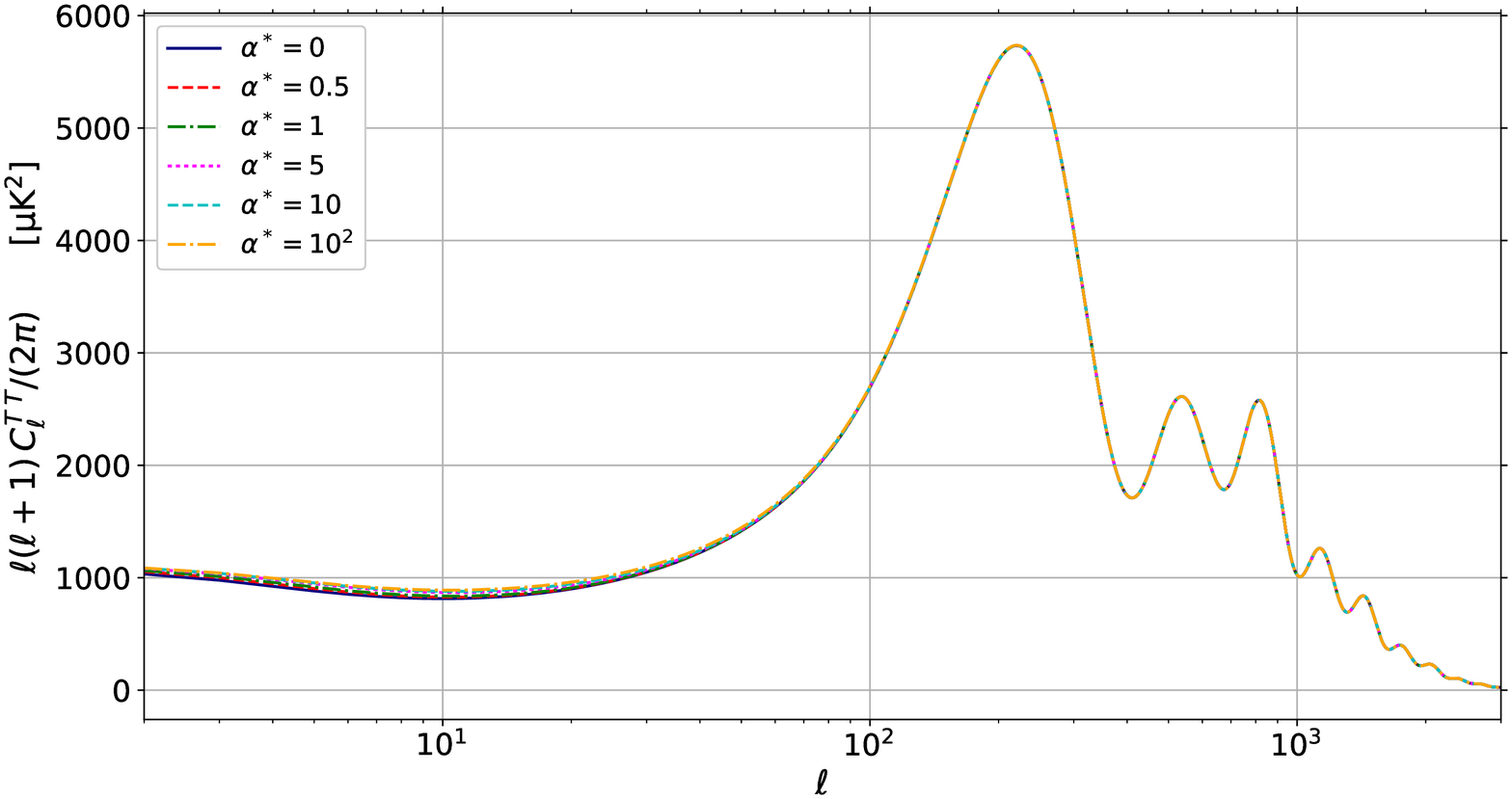} 
	\includegraphics[width=8cm]{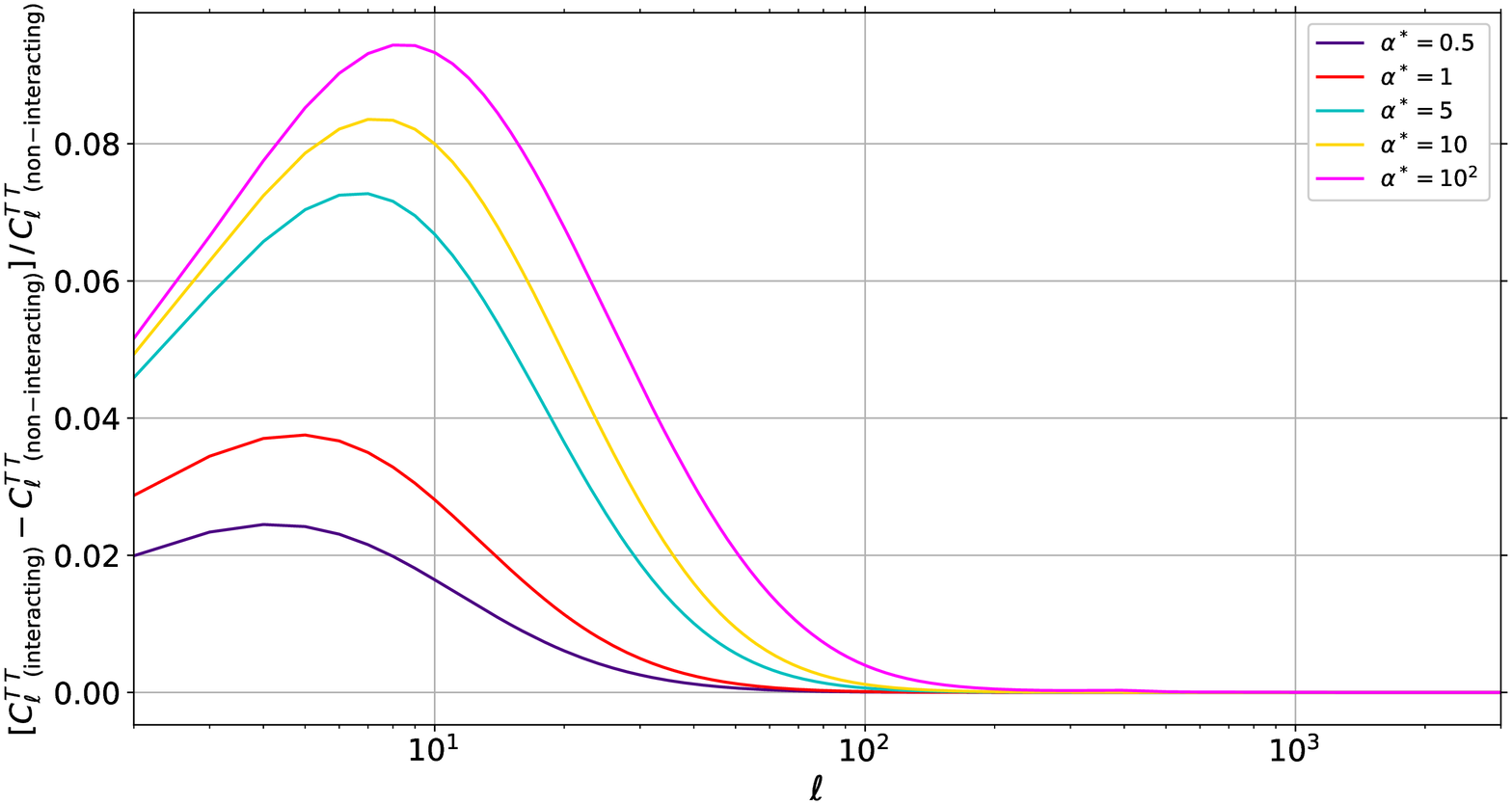} 
		\includegraphics[width=8cm]{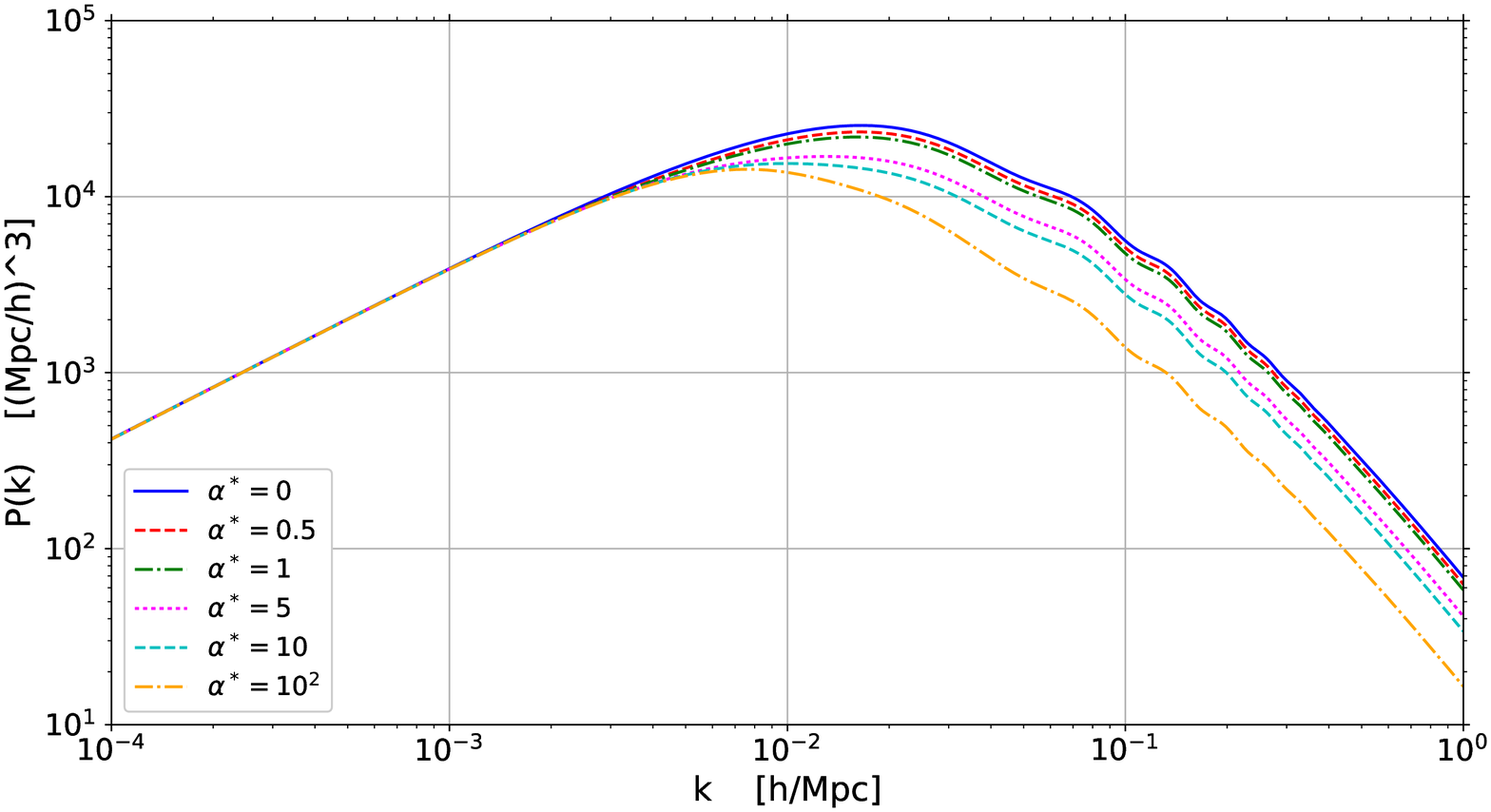} 
	\includegraphics[width=8cm]{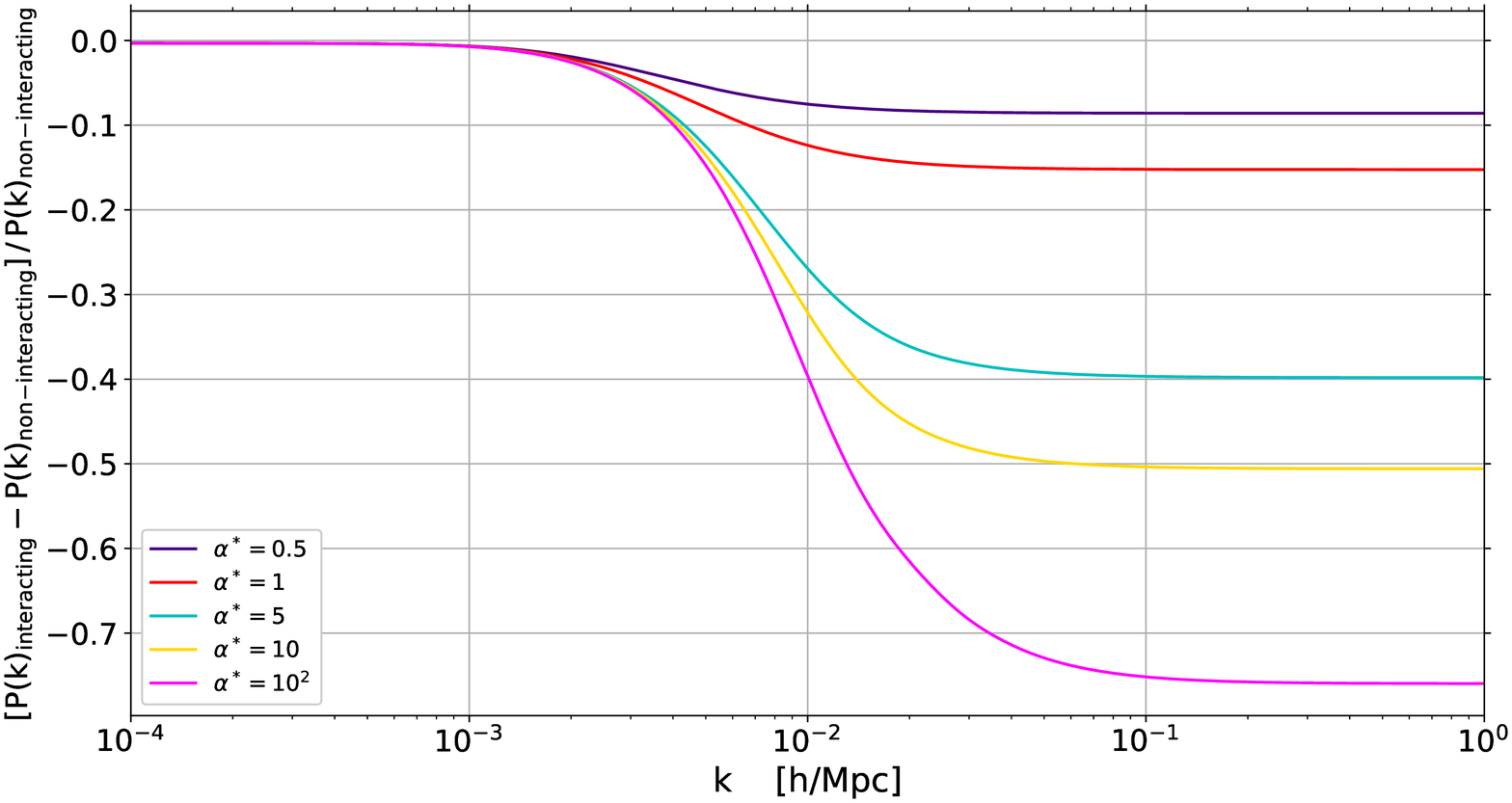}  
	\caption{The upper panels show the CMB power spectra (left) and their relative ratio (right) with respect to the $\Lambda$CDM model for different values of the interaction parameter. The lower panels show analogous plots for the matter power spectrum.}
	\label{cl}
\end{figure}

In Fig. \ref{cl} we show the CMB and matter power spectra for different values of the interaction parameter. We can see that the CMB is only modified on large scales, as expected because the interaction is only active at recent times and affects the CMB through a late time Integrated Sachs-Wolfe effect. The interaction affects more notoriously the matter power spectrum because the effect is more direct and not only through the modified gravitational potential. In the lower panels of Fig. \ref{cl} we can see that the very large scales beyond the horizon today are not affected by the interaction, again as one would expect because the interaction is designed as to only affect sub-horizon scales. As we go to smaller scales two distinctive regimes can be identified. On sufficiently small scales, there is suppression of the matter power spectrum that does not depend on $k$, being the suppression more important for larger values of the interaction parameter. At intermediate scales, there is a $k$-dependent suppression that interpolates between the unaffected large scales and the $k$-independent small scales suppression. This is easy to understand from the coupling to DE because we are assuming $c^2_{s\mathrm{(DE)}}=1$, thus dragging DM, smoothing out its clustering and, consequently, leading to a lower growth of structures. A similar effect can also be achieved in models where DM is coupled to a thermal bath of dark radiation which erases structures in the DM distribution (see e.g. \cite{Cyr-Racine:2013fsa,Buen-Abad:2015ova,Lesgourgues:2015wza,Ko:2016uft,Chacko:2016kgg,Ko:2017uyb,Buen-Abad:2017gxg,Raveri:2017jto}).

A very remarkable feature in the matter power spectrum that is very characteristic of this interacting model is the shift in the turnover. This peak is generally determined by the horizon at equality of matter and radiation because it simply accounts for the less rapidly growth of the density contrast of sub-Hubble modes during radiation as compared to the corresponding growth in the matter epoch (Meszaros effect). Since our interaction does not affect the background, one may expect the turnover to occur at the same equality scale. However, as we discussed above and can be explicitly seen in Fig \ref{velocities}, once the interaction term becomes relevant, the DM-DE system behaves as a single fluid and the growth of structures freezes, similarly to what happens during the radiation era due to the Meszaros effect. This happens for the band of modes that entered after the time when the interaction is turned on and it results in the shift in the position of the turnover that we see. This is a very unique signature of this model that is very hard to replicate with other models, which usually require modifying $\Omega_{\DM}$ to shift the turnover. However, this shift towards larger scales of the matter power spectrum maximum could be more prominently hidden by biasing effects \cite{BeltranJimenez:2010bb}.

Given that the power spectrum is suppressed by the interaction on sub-Hubble scales, the value of $\sigma_8$ can be reduced without modifying the background cosmology, i.e., with the same amount of DM as the standard $\Lambda$CDM model. As we will see in the next section, this remarkable feature will indeed help alleviating the $\sigma_8$ tension.

In order to gain a better understanding of the interacting model and how the effects on the CMB and matter power spectra are generated, we will now analyse the evolution of the peculiar velocities and density contrasts. This will also allow us to confirm our analytical findings above.

%%%%%%%%%%%%%%%%%%%%%%%%%%%%%%%%%%%%%%%%%%%%%%%%%%%%%%%%%%%
%%%%%%%%%%%%%%%%%%%%%%%%%%%%%%%%%%%%%%%%%%%%%%%%%%%%%%%%%%%
\subsection{Peculiar velocities} \label{sec4.2}

\begin{figure}[ht!]
    	\includegraphics[width=8cm]{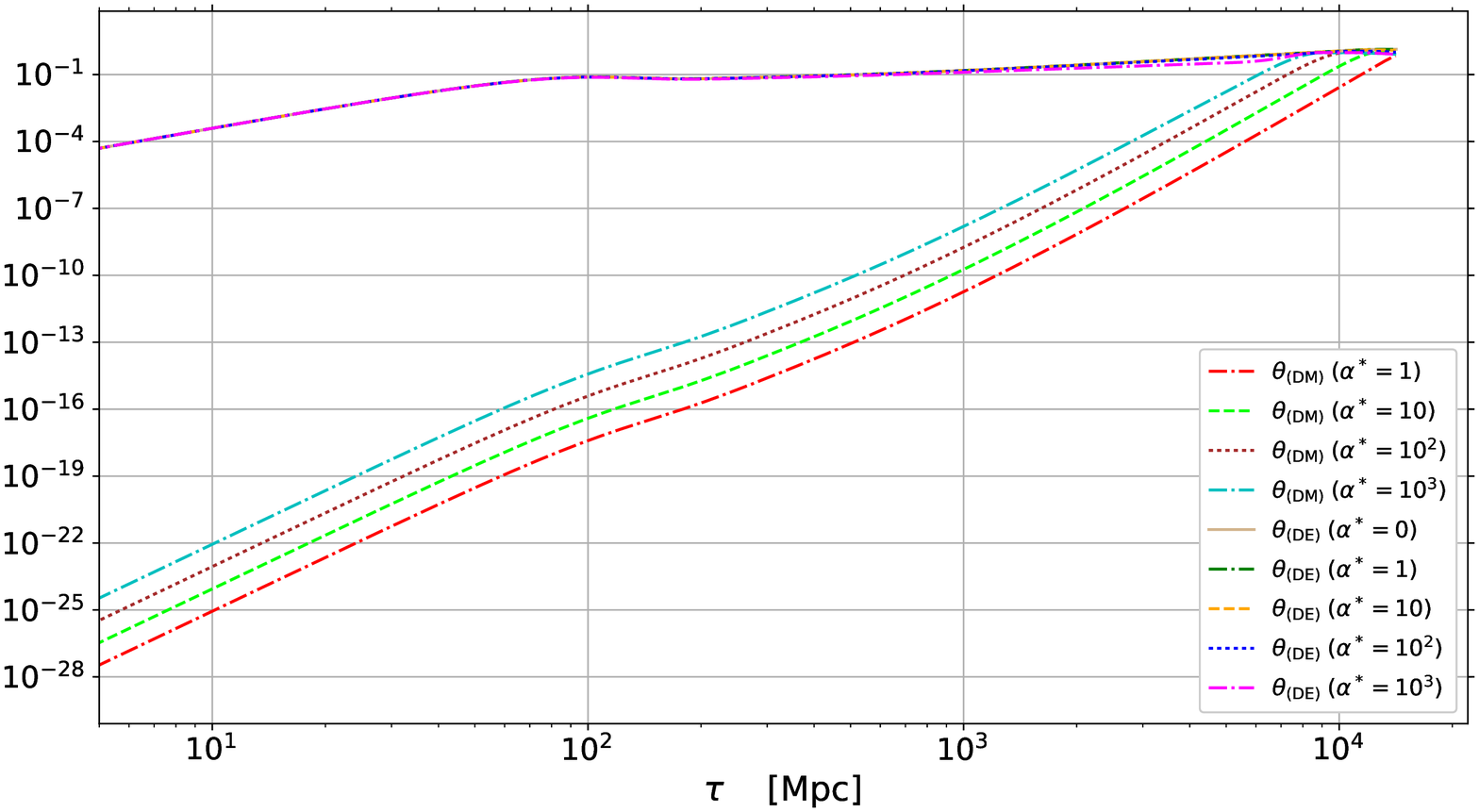}  
	    \includegraphics[width=8cm]{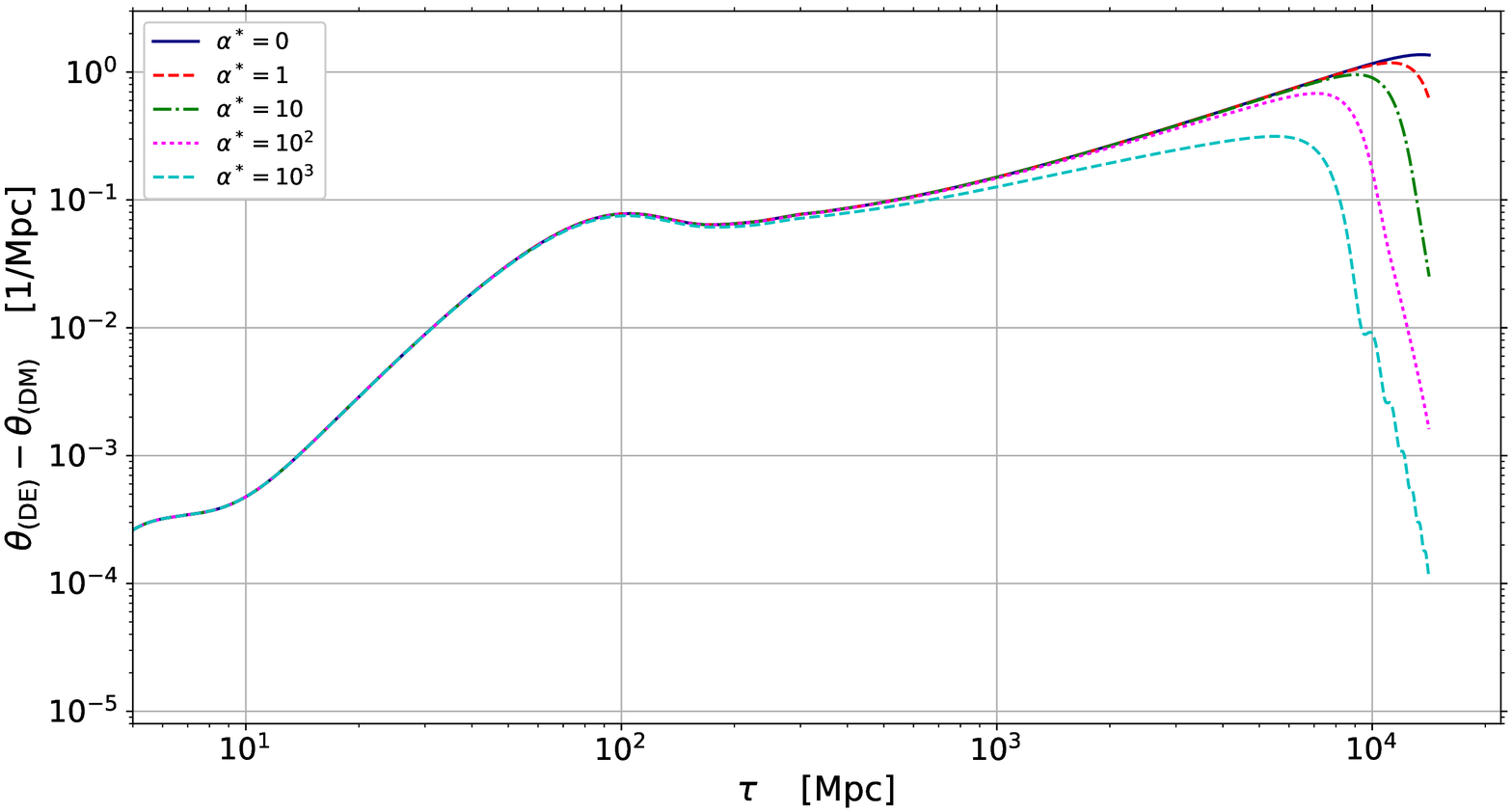} 
	    \includegraphics[width=8cm]{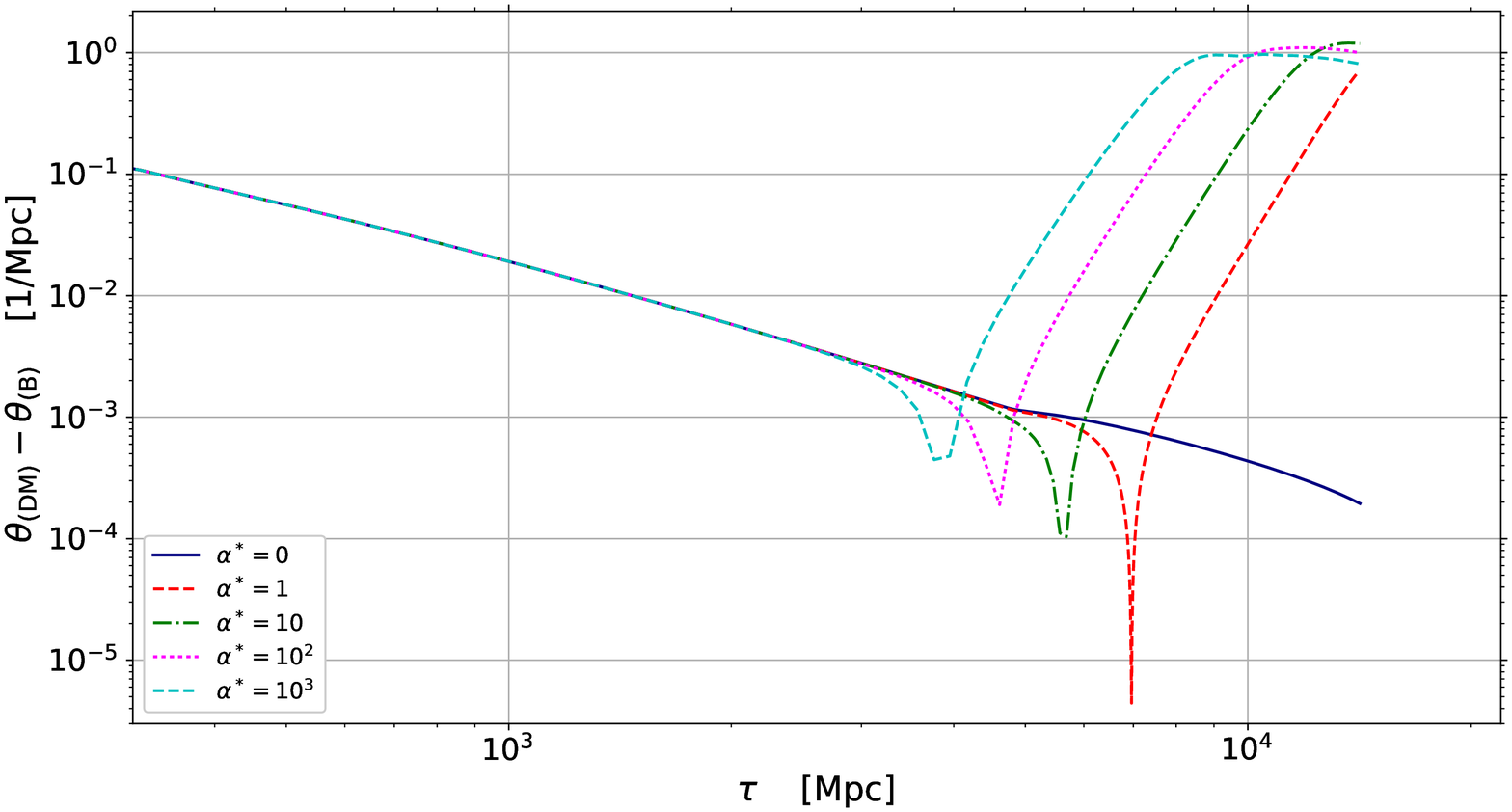}
	    \includegraphics[width=8cm]{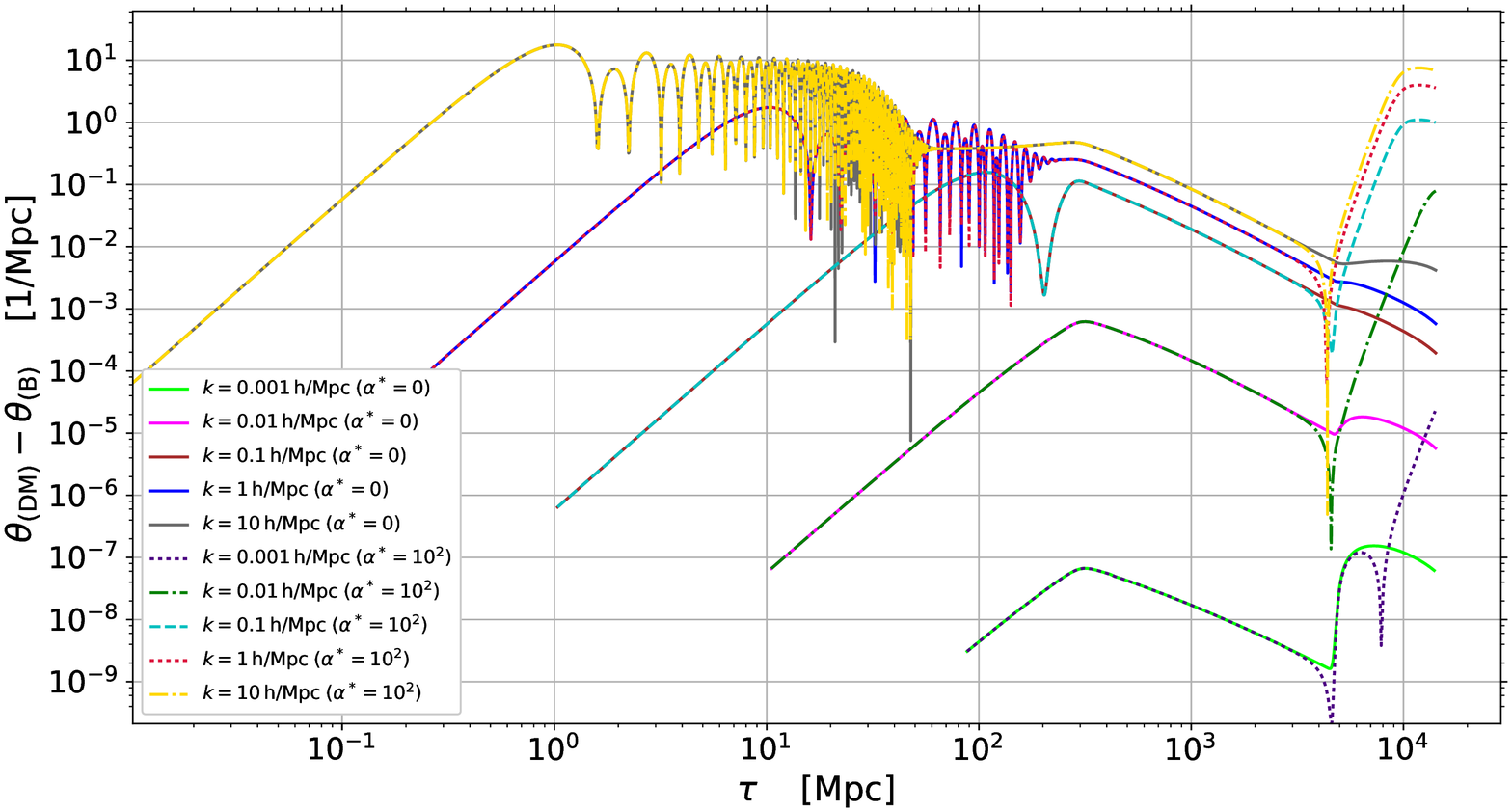}
	    \includegraphics[width=8cm]{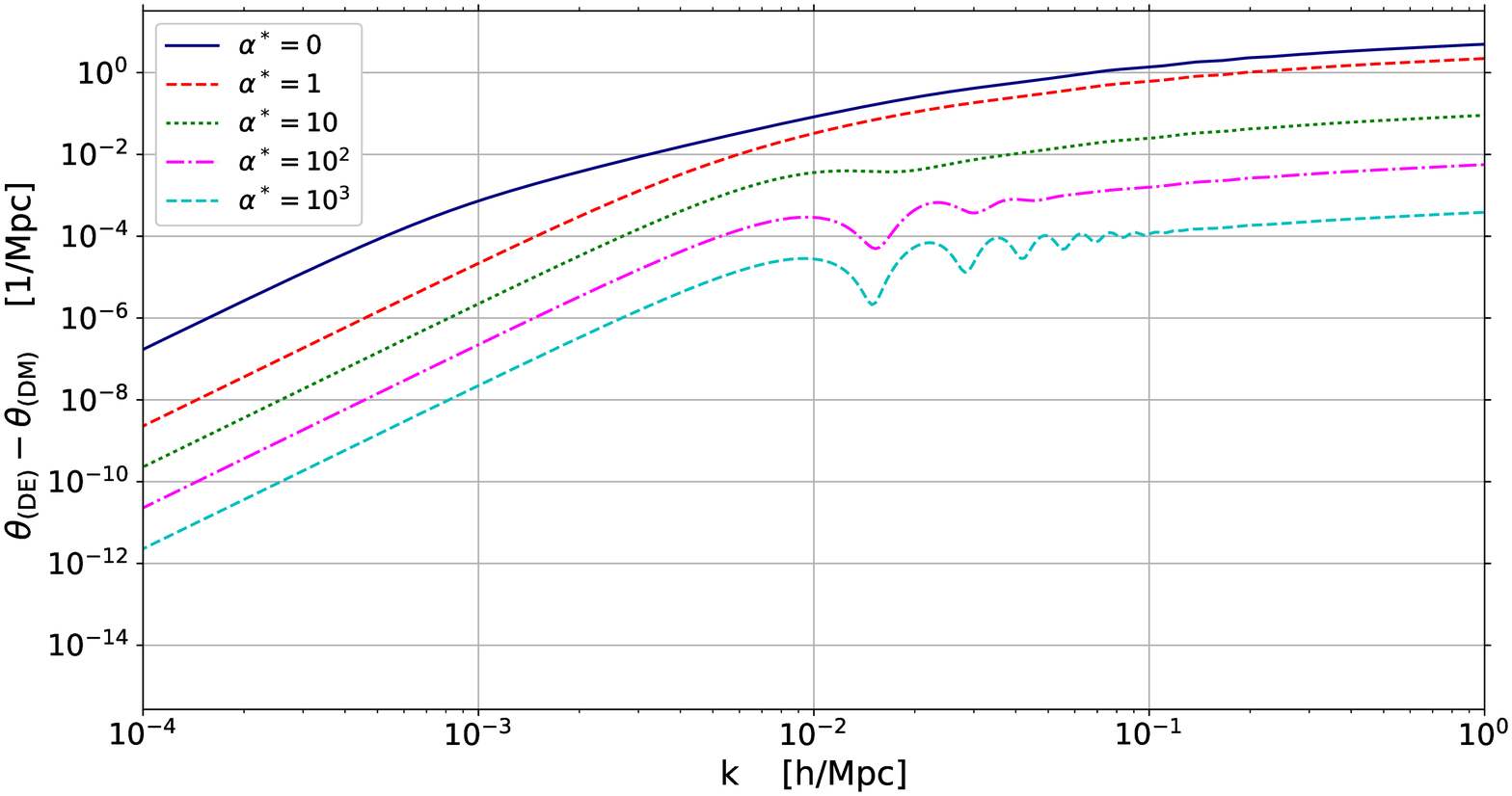}
	    \includegraphics[width=8cm]{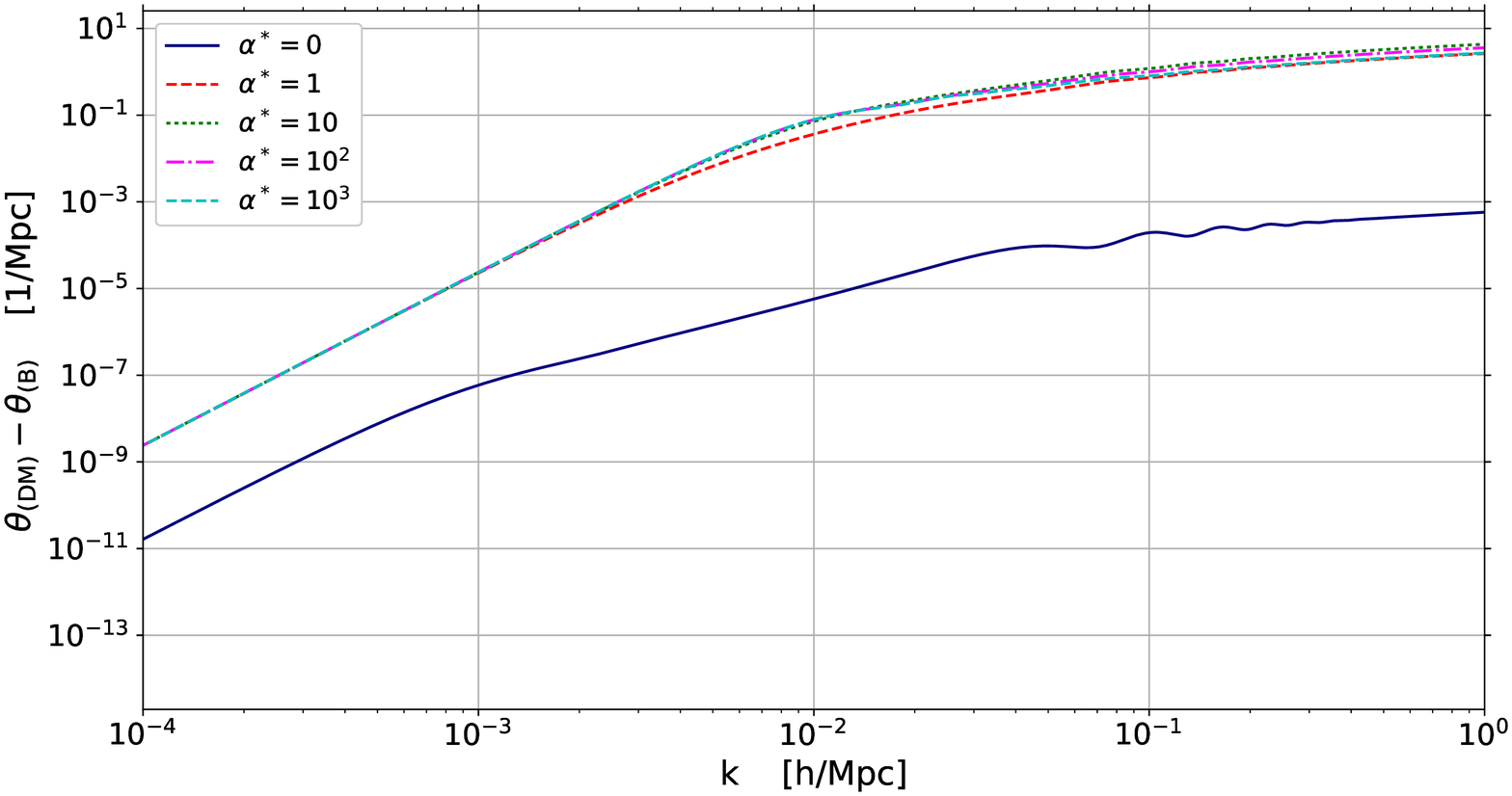}
	\caption{This figure shows the evolution and the spectrum of the velocity perturbations. The upper panels show the DM and DE velocities of one Fourier mode with $k=0.1\, h$/Mpc in synchronous gauge (left) and their relative velocity (right). The middle left panel shows the relative velocity of DM and baryons for the same Fourier mode, while the middle right panel shows the DM-baryons relative velocity in different scales. Finally, the bottom panels show the relative velocities today between DM and DE (bottom-left) and DM and baryons (bottom-right) as a function of the scale $k$.}
    \label{velocities}
\end{figure}

Since the interaction is directly given by the relative velocity of dark matter and dark energy, it will be convenient to analyse the evolution of the peculiar velocities. Fig. \ref{velocities} shows the evolution of the velocity perturbations of the different components. Let us recall here that the relative velocity that drives the interaction is gauge independent.

In the top-left panel we see the evolution of the peculiar velocities of DM and DE in synchronous gauge. We can see the large hierarchy between the DM and DE velocities in this gauge as obtained in (\ref{eq35}) and which is induced by the hierarchy between $\HH$ and the interaction $\zeta$ in the early universe. In this plot we can also confirm our analytical estimates in (\ref{eq33}) and (\ref{eq34}) for the growth of the peculiar velocities as $\theta_{\DE}\propto \tau^3$ and $\theta_{\DM}\propto \tau^8$ for DE and DM respectively. In fact, this large hierarchy is responsible for part of the $k$-dependent suppression on the matter power spectrum. One would naively expect a $k$-dependent effect for modes that enter between the time when the interaction is efficient, i.e. when $\zeta\sim \HH$ and today. However, we see that the $k$-dependent suppression in the matter power spectrum extends to much smaller scales. The reason can be understood as follows in the synchronous gauge: the interaction will tend to lock the DM-DE system so they share the same velocity, but, since the DM velocity started with a value much smaller than the DE velocity, there will be a certain delay for modes that did not enter early enough as to have been able to grow sufficiently by the time when the interaction is turned on. Since very sub-Hubble modes have been growing for a longer time, they are able to catch up with the DE velocity and this effect is no longer there, thus giving the approximately $k$-independent suppression in the matter power spectrum on very small scales that we see in Fig. \ref{cl}.

An interesting effect of the interaction between DM and DE is the appearance of a relative velocity between baryons and DM on small scales\footnote{This feature resembles the relative motions that appear among all the cosmic components in models with moving DE \cite{Maroto:2005kc,BeltranJimenez:2007rsj,Jimenez:2008vs,Harko:2013wsa}. However, while the relative motions appear on all scales for moving DE models, the interaction considered in this work only generates appreciable relative motions between baryons and DM on sub-horizon scales.}. We show this feature in Fig. \ref{velocities} where we can see how the relative velocity between baryons and DM grows once the interaction is turned on (middle panels). An important consequence of this effect will be a contribution to the bias parameter and an off-center location of galaxies inside DM haloes. In the lower panels we show the relative velocity today of baryons (bottom-right) and DE (bottom-left) with respect to DM as a function of the scale $k$. The relative velocity of DE and DM (bottom-left) shows some acoustic oscillations on scales between $k=10^{-2}$ - $10^{-1}h$/Mpc that are caused by the coupling of the pressureless DM component and DE with $c_{s\mathrm{(DE)}}=1$ that gives rise to acoustic oscillations very much like in the baryon-photon system before recombination. Similar oscillations in the DM sector are also produced when DM is coupled to a thermal background of dark radiation as in e.g. \cite{Cyr-Racine:2013fsa,Chacko:2016kgg,Buen-Abad:2017gxg,Raveri:2017jto}. The analogous role of the dark radiation is played here by DE. 

Concerning the DM-baryons relative velocity, it is worth explaining that, in the absence of interactions, it is possible to go to the DM comoving gauge so the relative velocity is dominated by that of baryons. Thus, the oscillations we see in the curve for $\alpha^*=0$ are in fact the usual baryon acoustic oscillations imprinted at recombination. When the interaction is switched on, the DM comoving gauge is no longer permitted and the DM velocity (in synchronous gauge) then approaches the attractor solution (\ref{eq34}) on super-Hubble scales in the radiation dominated epoch. This explains the mismatch between the non-interacting (blue-solid curve) and the interacting (remaining non-solid curves) cases seen on large scales in the bottom-right panel of Fig. \ref{velocities}. For the interacting cases, all the curves coincide on large scales as expected and the relative velocities grow as we go to smaller scales. The growth is not very sensitive to the value of the coupling parameter $\alpha^*$ and we see that the velocities can become substantially appreciable on small scales, what will lead to the aforementioned bias between galaxies and DM haloes.

%%%%%%%%%%%%%%%%%%%%%%%%%%%%%%%%%%%%%%%%%%%%%%%%%%%%%%%%%%%%
%%%%%%%%%%%%%%%%%%%%%%%%%%%%%%%%%%%%%%%%%%%%%%%%%%%%%%%%%%%%
\subsection{Density perturbations and Newtonian potential} \label{sec4.3}

\begin{figure}[hb!]
   %\centering
	\includegraphics[width=8cm]{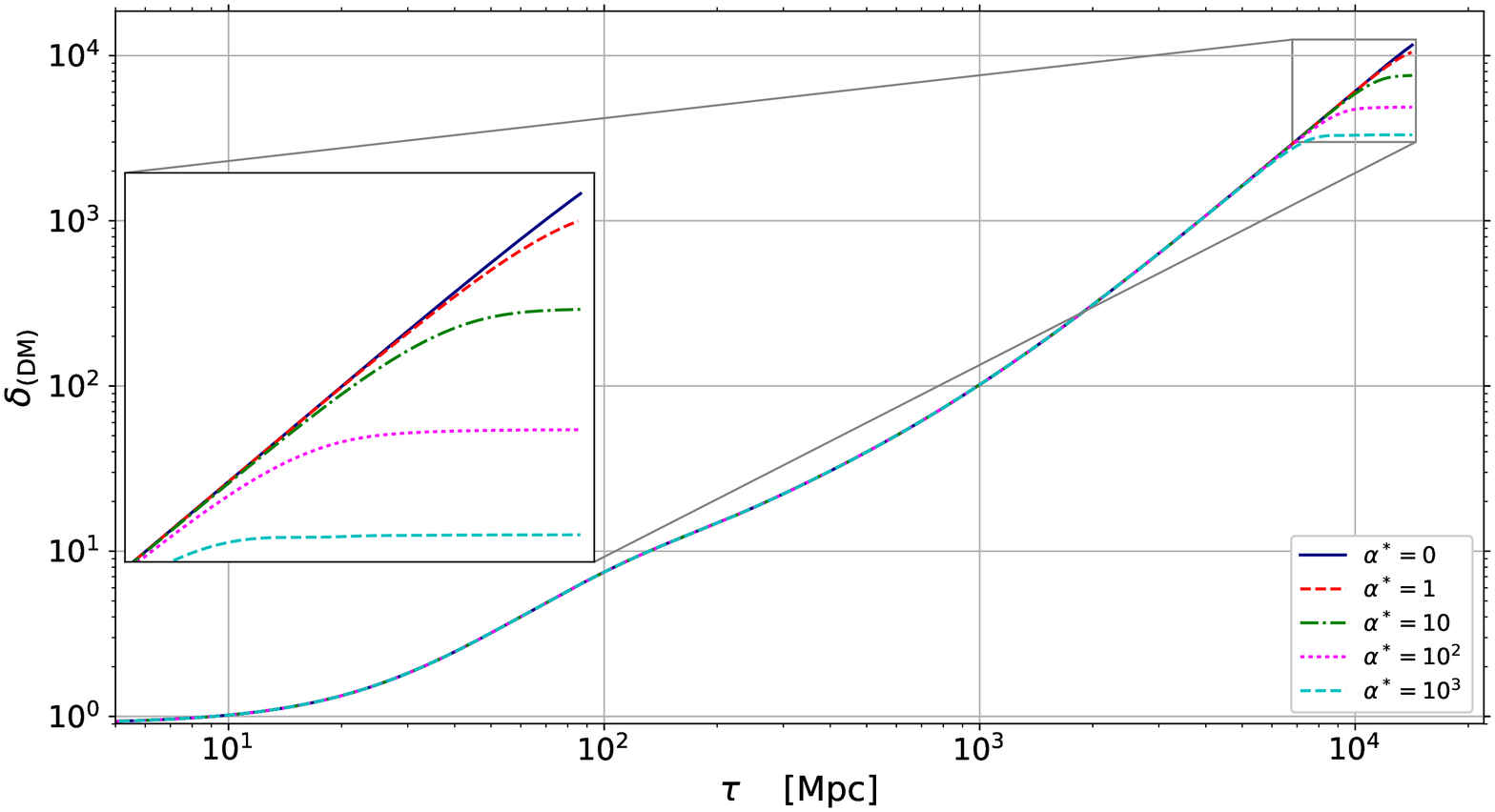}  
	\includegraphics[width=8cm]{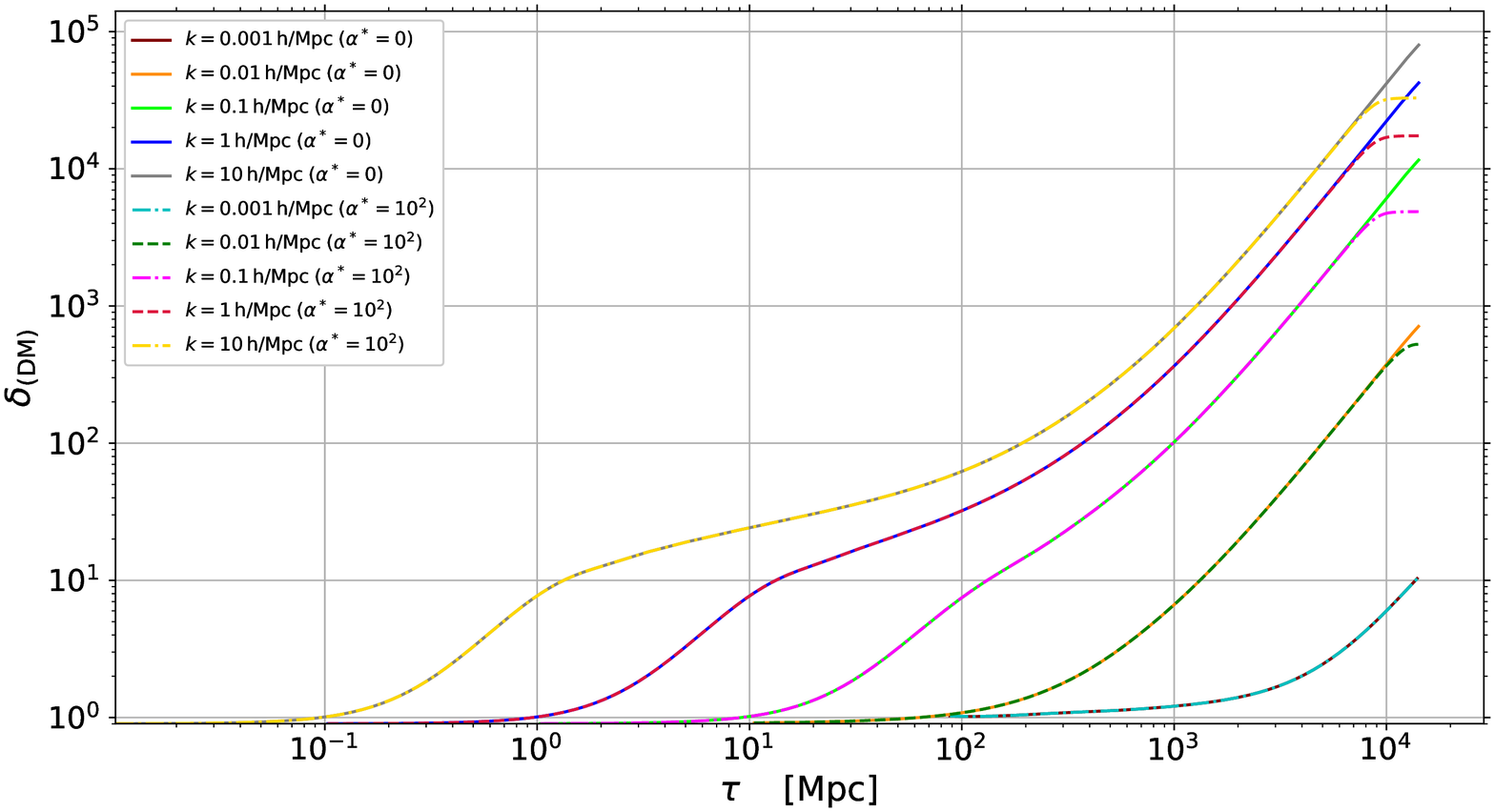}  
	\includegraphics[width=8cm]{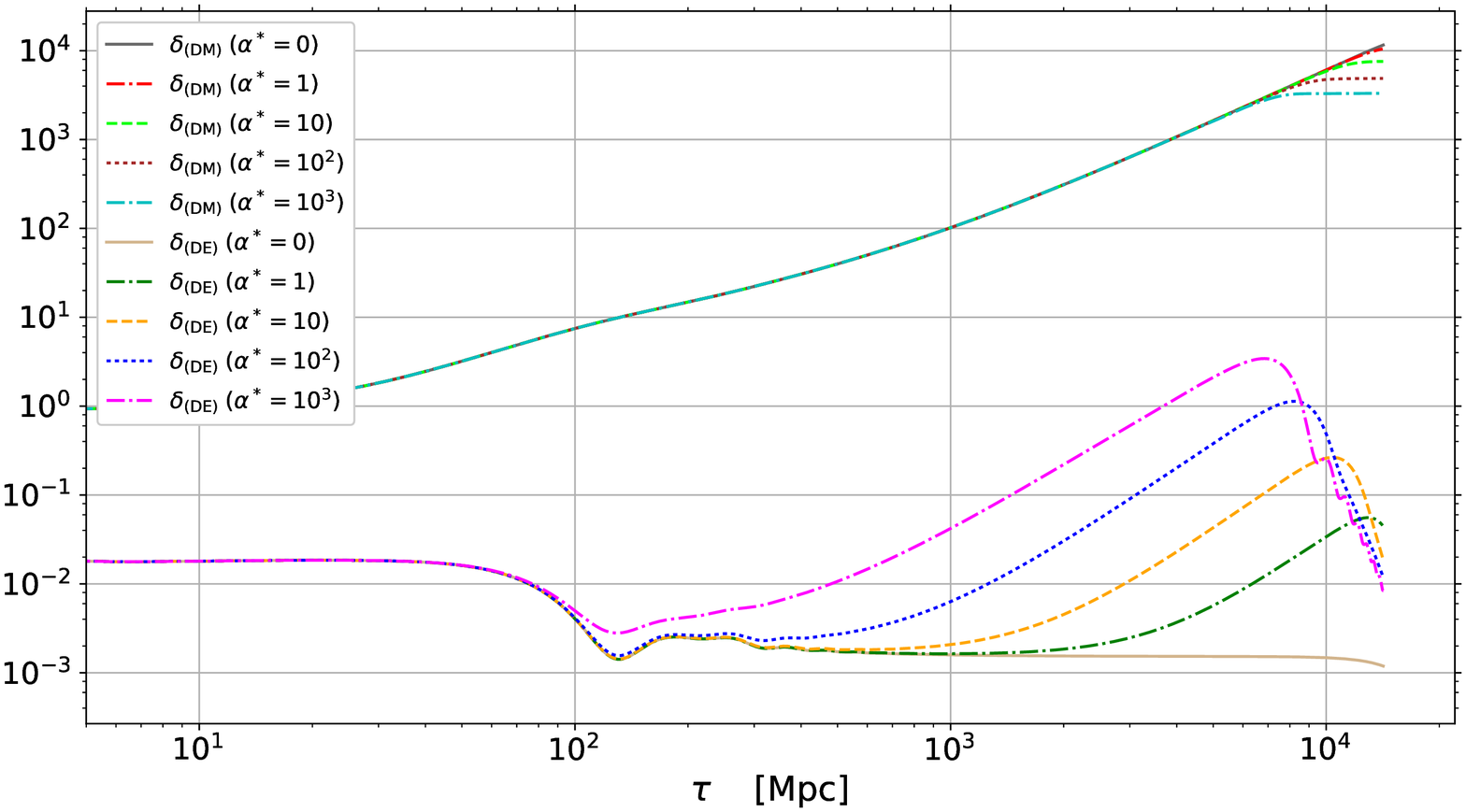}  
	\includegraphics[width=8cm]{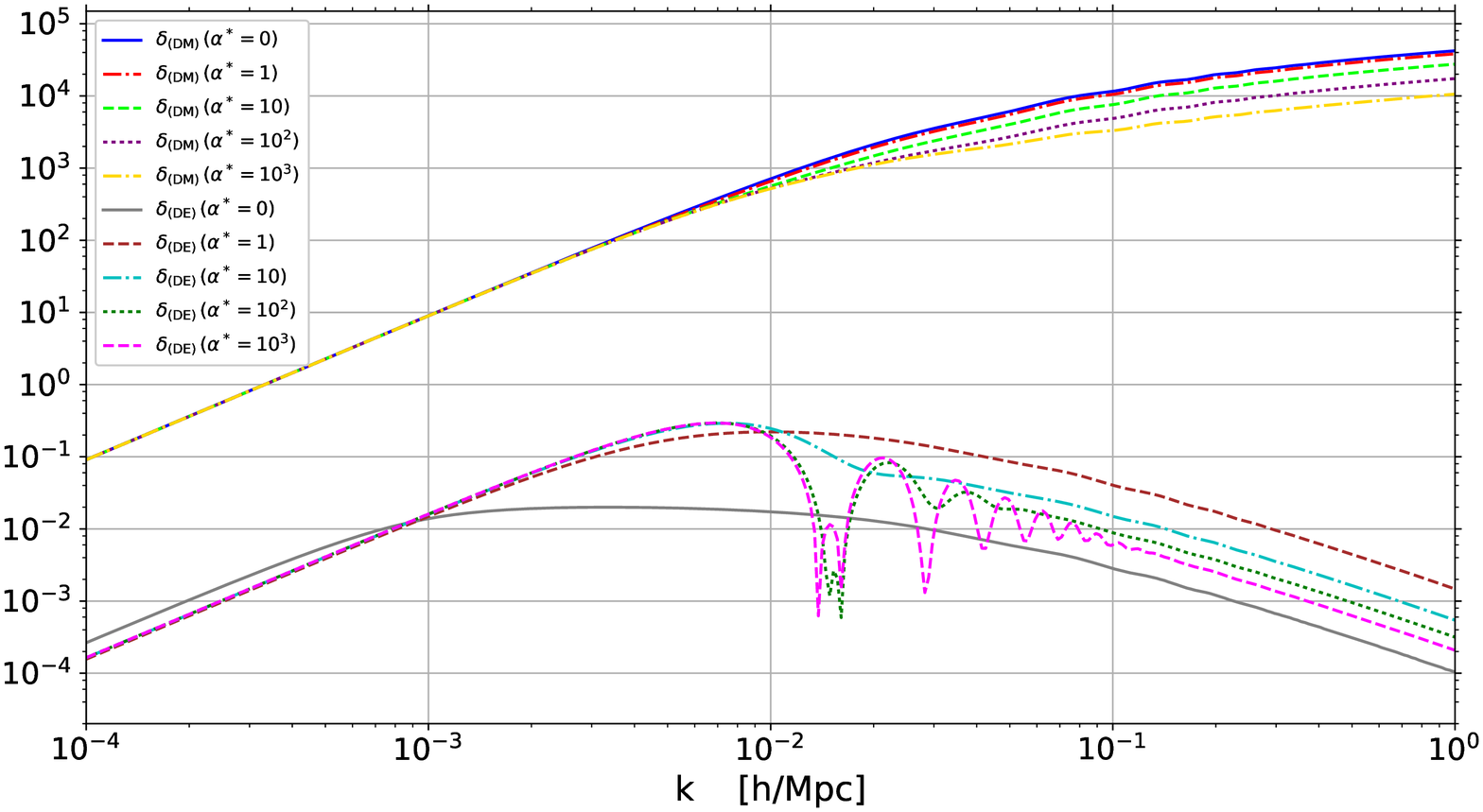} 
	\includegraphics[width=8cm]{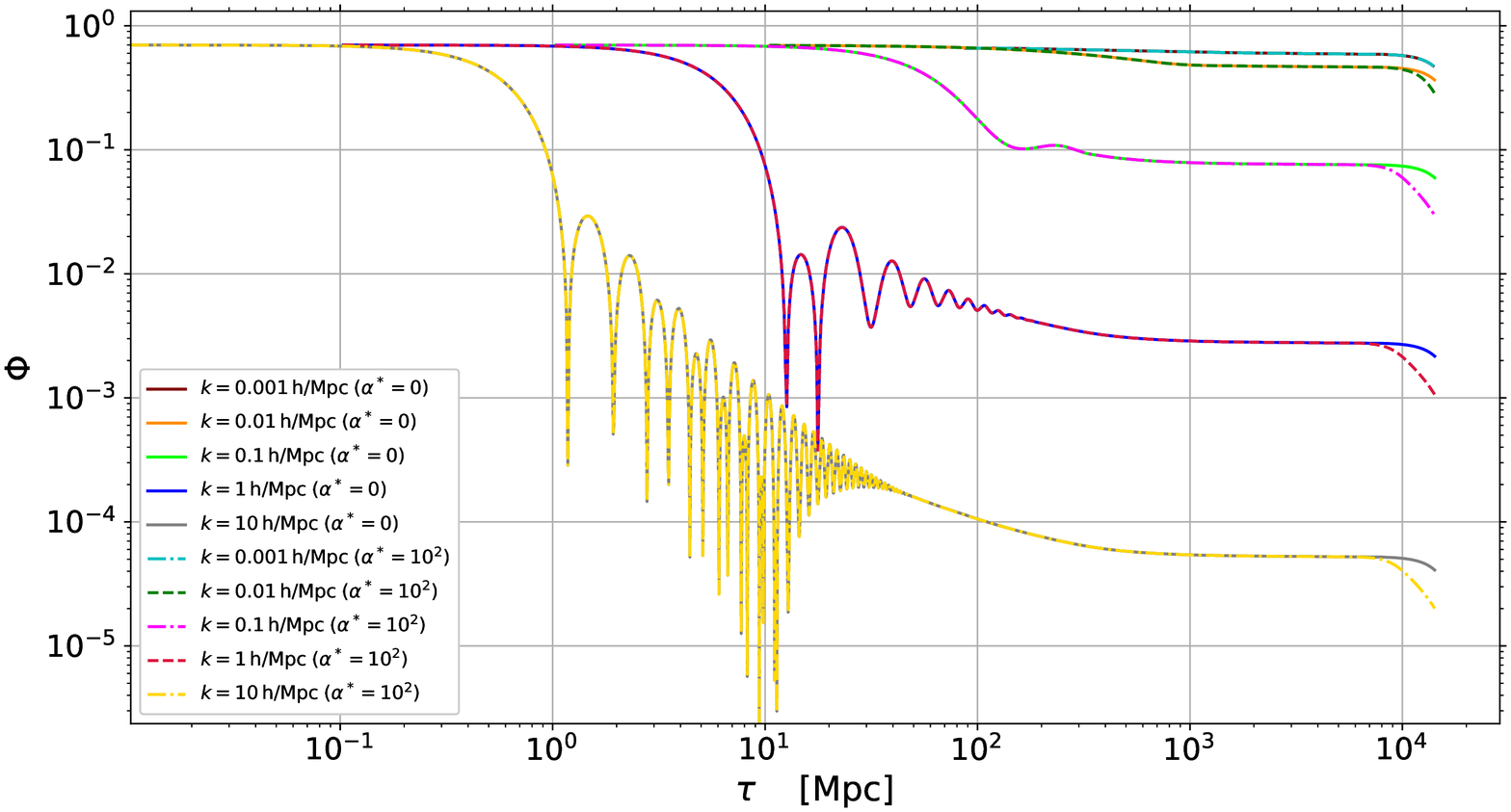}  
	\includegraphics[width=8cm]{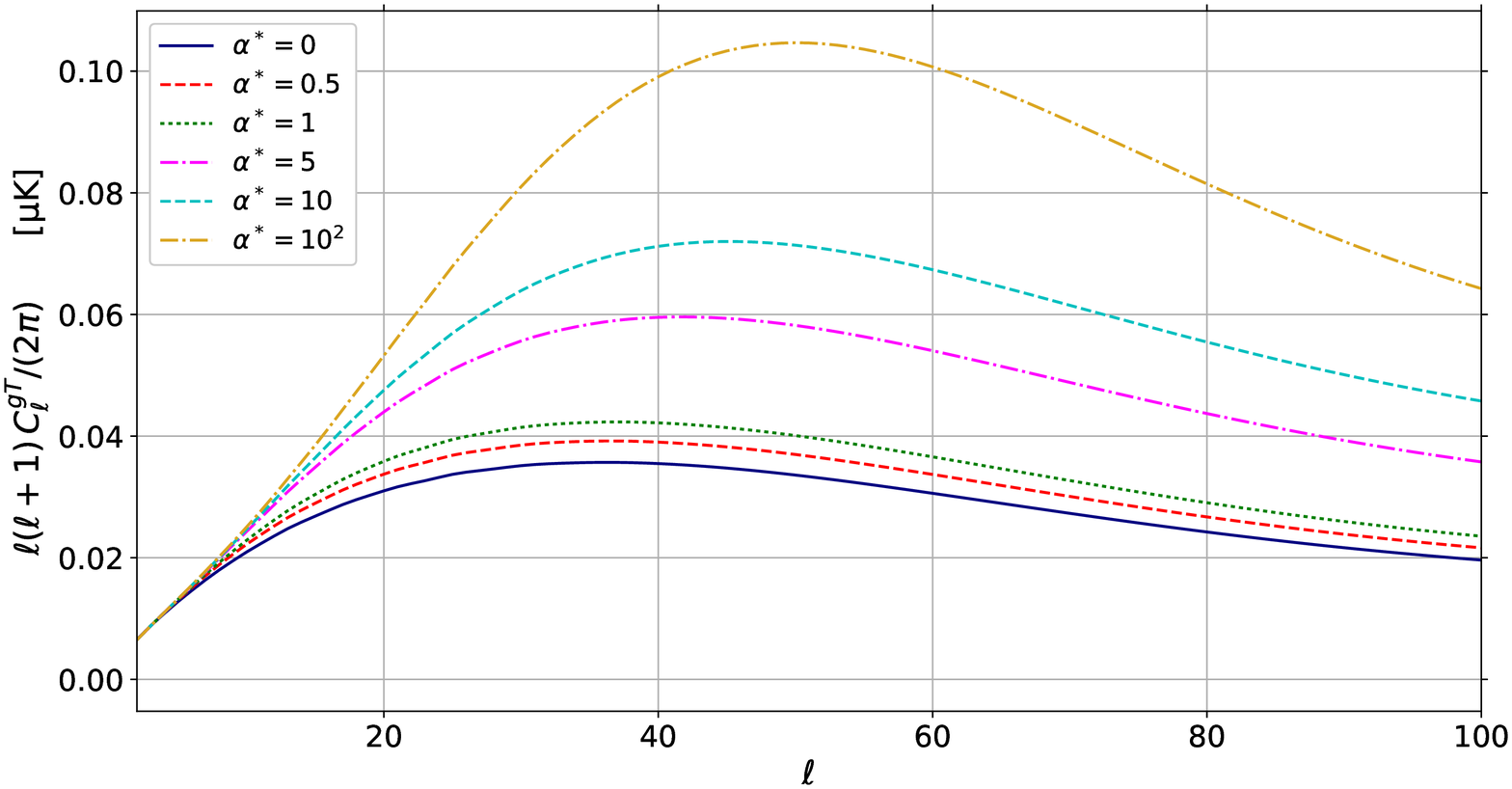}  
	\caption{This Figure shows the evolution of the DM density contrast for different values of the interaction parameter  and fixed Fourier mode with $k=0.1\,h$/Mpc (upper-left) and for different Fourier modes (upper-right). The middle-left panel shows the comparison with the DE density contrast evolution, where we see that the DE density contrast is very sub-dominant with respect to the DM density contrast. This is confirmed in the middle-right panel where we show the density perturbations today as a function of the scale $k$. In the bottom-left panel we show the evolution of the Newtonian potential and the bottom-right panel shows the CMB-galaxies cross-correlation. We see the more rapid decay of the Newtonian potentials at late-times, what leads to the positive contribution to the ISW, thus enhancing the CMB-galaxies cross-correlation.}
	\label{densities}
\end{figure}

We now turn our attention to the evolution of DM and DE density perturbations that will eventually determine the growth of structures and that we show in Fig. \ref{densities}. In the upper-left panel we show the evolution of the DM density contrast (in conformal gauge) for a given Fourier mode ($k=0.1\,h$/Mpc) and different strengths of the interaction. We see that the evolution is the same as in $\Lambda$CDM until $\tau_\zeta$. After this time, both components are tightly coupled, they will be dragged by each other so that their relative velocity drops (see Fig. \ref{velocities}) and this further implies, according to Eq. (\ref{eq39}), that the density contrast of DM freezes. This is the phenomenon that we observe in the numerical solutions shown in Fig. \ref{densities}. As one would expect, the growth of structures ceases earlier for larger values of the interaction constant. 

The middle-left panel of Fig. \ref{densities} shows the evolution of the DE density contrast (in conformal gauge) where we can confirm that the density contrast is much smaller than the DM density contrast, as one would expect from imposing that DE has $c^2_{s\mathrm{(DE)}}=1$. After the interaction becomes efficient, we thus have a system that resembles the baryon-photon plasma before recombination. One would then may expect to see acoustic oscillations very much like the BAO. This is in fact the case, but the amplitude of the oscillations in the DM density contrast for the DM-DE coupled system is very small as to be seen for the parameter range considered here. These oscillations are more apparent in the evolution of the velocity perturbations in Fig. \ref{velocities} and in the density contrast of DE (left-middle panel of Fig. \ref{densities}). In the middle-right panel we show the density contrasts of DM and DE today as a function of the scale $k$. In that plot we can again confirm that the DE density perturbations are small as compared to those of DM on all scales. Furthermore, we can see again that the acoustic oscillations in the DM-DE system appear more prominently in the DE sector, while in the DM are not visible for our choice of parameters. We see once more the suppression of the growth of structures on small scales for DM as compared to $\Lambda$CDM for the same set of cosmological parameters except for the strength of the interaction.

In the bottom-left panel of Fig. \ref{densities} we plot the evolution of the Newtonian potential. We can see how once the interaction becomes efficient, there is a suppression in the Newtonian potential. This suppression induces a time-variation of the Newtonian potential in addition to the one due to entering the DE domination epoch so that the interaction will give a contribution to the late-time Integrated Sachs-Wolfe effect, as we discussed above. The contribution of this effect to the total CMB power spectrum is negligible, but it can be more efficiently detected by cross-correlating the CMB with large scale structures or, more precisely, with galaxy surveys (see e.g. \cite{Giannantonio:2008zi,Ferraro:2014msa,Ade:2015dva}). In the bottom-right panel of Fig. \ref{densities} we show the cross-correlation between CMB and galaxies for the standard $\Lambda$CDM and the interacting model. We see how the signal is clearly enhanced as we increase the strength of the interaction and this enhancements is towards more positive correlations. We will not use ISW-LSS data to constraint the interacting model because that requires a careful account of the bias due to the existing degeneracy. For our interacting model, there is the additional contribution to the bias arising from the larger relative between baryons and DM that we have discussed above. However, it is clear that this observable has the potential to probe the type of interactions under consideration in this work.

%%%%%%%%%%%%%%%%%%%%%%%%%%%%%%%%%%%%%%%%%%%%%%%%%%%%%%%%%%%%%%
%%%%%%%%%%%%%%%%%%%%%%%%%%%%%%%%%%%%%%%%%%%%%%%%%%%%%%%%%%%%%%%
\section{Fit to observational data} \label{sec4.4} 
After analysing the most important features of the evolution of the perturbations for the interacting model under consideration in this work, we will now proceed to compare it with cosmological observations. For that, we will run the Markov chain Monte Carlo code M\textsc{onte} P\textsc{ython} \cite{Audren:2012wb,Brinckmann:2018cvx}, in order to obtain the corresponding constraints on the parameters of the interacting model. We will consider the following set of parameters in the MCMC analysis:
\begin{equation}
	\left\{ 
	100\,\Omega_{\mathrm{(B)},0} h^2,\, 
	\Omega_{\mathrm{(DM)},0} h^2,\, 
	100\,\theta_s,\, 
	\ln (10^{10} A_s),\, 
	n_s,\, 
	\tau_{\mathrm{reio}},\, 
	w_{\mathrm{(DE)}},\, 
	\alpha^* \right\} 
\end{equation}
where $\Omega_{\mathrm{(B)},0} h^2$ and $\Omega_{\mathrm{(DM)},0} h^2$ are the baryon and cold dark matter densities relative to the critical density respectively, $\theta_s$ is the ratio of the sound horizon to the angular diameter distance at decoupling, $A_s$ is the amplitude of the primordial scalar perturbation spectrum, $n_s$ is the scalar spectral index, $\tau_{\mathrm{reio}}$ is the optical depth to reionization, $w_{\mathrm{(DE)}}$ is the dark energy equation of state parameter. The additional parameter with respect to the standard model that we introduce is the parameter $\alpha^*$ that determines the strength of the interaction. In principle, we do not have any prior knowledge on this parameter, but the preliminary numerical work suggests that a flat prior in the range [$-0.1$, $100$] is a sensible choice. Besides the primary parameters to be varied, we have four derived important parameters, namely: the Hubble constant ($H_0$), the reionization redshift ($z_{\mathrm{reio}}$), the matter density parameter ($\Omega_{\mathrm{(M)},0}$), and the root-mean-square mass fluctuations on scales of 8 $h^{-1}$ Mpc ($\sigma_8$).  

In this analysis we use six likelihoods: The Planck likelihood containing high-$\ell$, low-$\ell$ and lensing data provided with the Planck Legacy Archive\footnote{\url{http://pla.esac.esa.int/pla/\#cosmology}} \cite{2016A&A...594A..13P}, the Planck-SZ likelihood for the Sunyaev-Zeldovich effect measured by Planck \cite{2016A&A...594A..24P,2014A&A...571A..20P}, the CFHTLenS likelihood with the weak lensing data \cite{doi:10.1093/mnras/stt041,doi:10.1093/mnras/stt601}, the JLA likelihood with the supernovae data \cite{2014A&A...568A..22B},  the BAO likelihood with the baryon acoustic oscillations data \cite{doi:10.1111/j.1365-2966.2011.19250.x,doi:10.1093/mnras/stu523}, and the BAORSD likelihood for BAO and redshift-space distortions (RSD) measurements \cite{doi:10.1093/mnras/stx721,Buen-Abad:2017gxg}. 

  \begin{table}[h!]
    	\centering
    	\scalebox{0.85}{
            \begin{tabular}{|c|c|c|c|c|} 
    	    \hline
    	     & \multicolumn{4}{|c|}{}  \\
   	         & \multicolumn{4}{|c|}{Planck + Planck-SZ + CFHTLenS + JLA + BAO + BAORSD}  \\
   	        \cline{2-5} 
   	         & \multicolumn{2}{|c|}{} & \multicolumn{2}{|c|}{} \\
   	         & \multicolumn{2}{|c|}{$\Lambda$CDM} & \multicolumn{2}{|c|}{Interacting model} \\
   	        \cline{2-5}  
      	     & & & & \\
      	    parameter &  Best fit & 68\% \& 95\% limits &  Best fit & 68\% \& 95\% limits \\ \hline 
      	     & & & & \\
      	    $100\,\Omega_{\mathrm{(B)},0}\,h^2$ & $2.260$ & $2.251^{+0.020+0.038}_{-0.019-0.038}$ & $2.232$ & $2.238^{+0.019+0.042}_{-0.021-0.040}$ \\
      	     & & & & \\
      	    $\Omega_{\mathrm{(DM)},0}\,h^2$ & $0.1153$ & $0.1153^{+0.0011+0.0020}_{-0.00087-0.0020}$ & $0.1182$ & $0.1171^{+0.0013+0.0028}_{-0.0014-0.0027}$ \\
      	     & & & & \\
      	    $100\,\theta_s$ & $1.042$ & $1.042^{+0.00038+0.00077}_{-0.00037-0.00075}$ & $1.042$ & $1.042^{+0.00039+0.00084}_{-0.00043-0.00082}$ \\
      	     & & & & \\
      	    $\ln (10^{10} A_s)$ & $3.036$ & $3.031^{+0.011+0.035}_{-0.022-0.030}$ & $3.068$ & $3.085^{+0.026+0.053}_{-0.025-0.053}$ \\
      	     & & & & \\
      	    $n_s$ & $0.9693$ & $0.9721^{+0.0037+0.0078}_{-0.0042-0.0078}$ & $0.9671$ & $0.9707^{+0.0048+0.010}_{-0.0051-0.0097}$ \\
      	     & & & & \\
      	    $\tau_{\mathrm{reio}}$ & $0.05684$ & $0.05374^{+0.0038+0.018}_{-0.014-0.014}$ & $0.07132$ & $0.07850^{+0.014+0.029}_{-0.014-0.028}$ \\
      	     & & & & \\
      	    $w_{\mathrm{(DE)}}$ & --- & --- & $-0.9987$ & $-0.9725^{+0.0070+0.049}_{-0.028-0.028}$ \\
      	     & & & & \\
      	    $\alpha^*$ & --- & --- & $1.230$ & $1.042^{+0.26+0.59}_{-0.31-0.53}$ \\
      	     & & & & \\
      	    $H_0\;\;[\mathrm{\frac{km}{s\,Mpc}}]$ & $69.86$ & $69.81^{+0.42+1.0}_{-0.52-0.92}$ & $68.51$ & $68.19^{+0.78+1.5}_{-0.67-1.5}$ \\
     	     & & & & \\
      	    $z_{\mathrm{reio}}$ & $7.813$ & $7.496^{+0.60+1.7}_{-1.2-1.5}$ & $9.323$ & $9.902^{+1.3+2.5}_{-1.1-2.5}$ \\
      	     & & & & \\
      	    $\Omega_{\mathrm{(M)},0}$ & $0.2825$ & $0.2828^{+0.0060+0.012}_{-0.0050-0.011}$ & $0.2995$ & $0.3000^{+0.0078+0.016}_{-0.0080-0.016}$ \\
      	     & & & & \\
      	    $\sigma_8$ & $0.8043$ & $0.8036^{+0.0047+0.013}_{-0.0069-0.011}$ & $0.7517$ & $0.7569^{+0.011+0.023}_{-0.012-0.022}$ \\
      	     & & & & \\
      	    \hline    
            \end{tabular}
        }    
        \caption{Best fit values and 68\% and 95\% confidence limits obtained from "Planck + Planck-SZ + CFHTLenS + JLA + BAO + BAORSD" data set for $\Lambda$CDM and our interacting model.}
        \label{tab:1}
    \end{table} 

  \begin{figure}[h!]
    	\centering
    	\scalebox{1.}{ 
        \includegraphics[width=\linewidth]{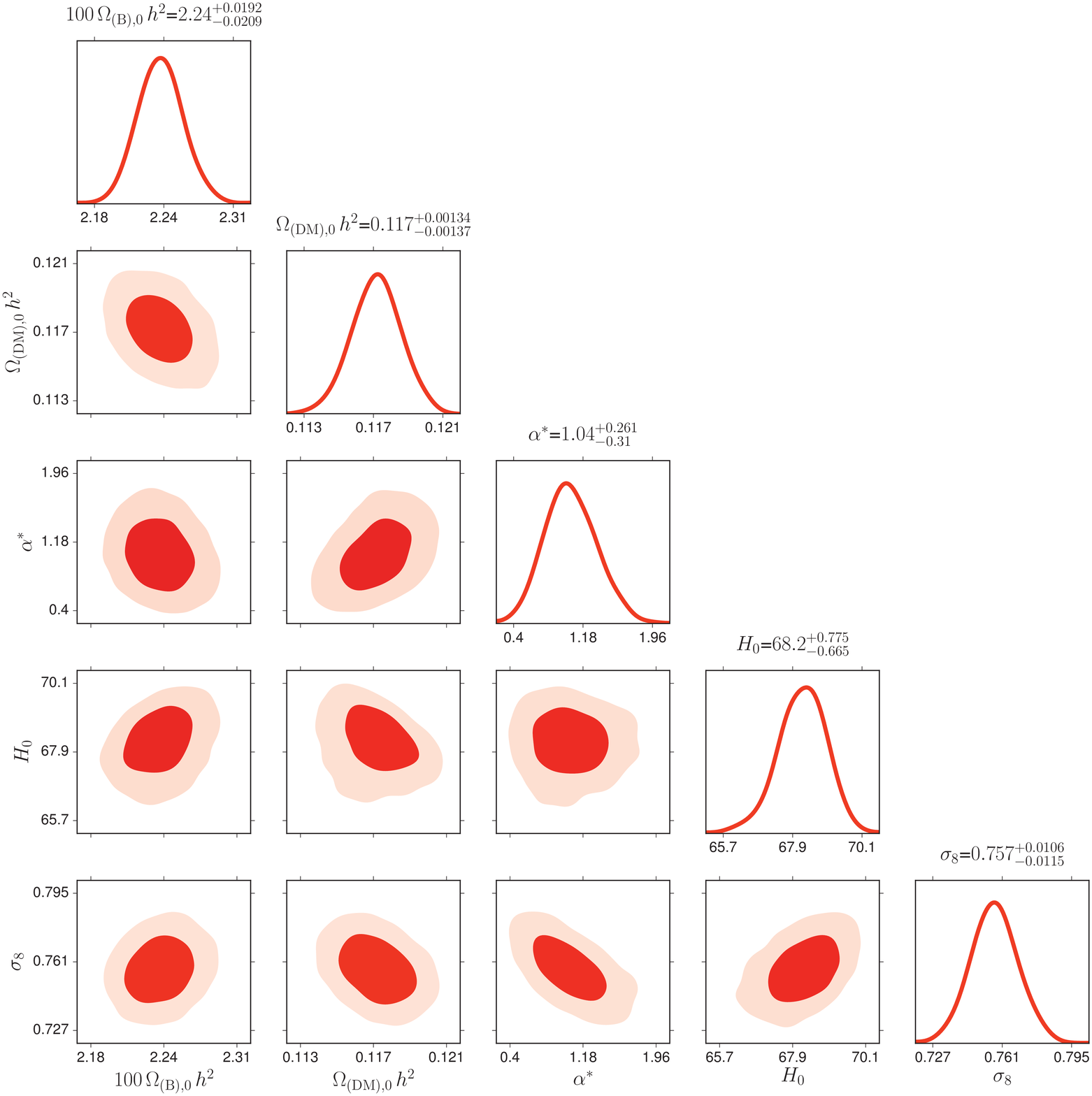} }
    	\caption{The one-dimensional posterior distribution and two-dimensional posterior contours with 68\% and 95\% confidence limits from the "Planck + Planck-SZ + CFHTLenS + JLA + BAO + BAORSD" data set, for the selected cosmological parameters of the interacting model.}
    	\label{fig:triangle}
    \end{figure}

The obtained constraints on the cosmological parameters, considering the combined "Planck + Planck-SZ + CFHTLenS + JLA + BAO + BAORSD" data set, are shown in Table \ref{tab:1}, for the standard $\Lambda$CDM and the interacting models.  The corresponding confident regions and marginalised posterior distributions are shown in Fig. \ref{fig:triangle}. We find two very remarkable results that should be highlighted. The first one is the lower value of $\sigma_8$ with respect to $\Lambda$CDM. This can be understood because, as we have seen above, the tight coupling of DM to DE after the interaction becomes efficient prevents the clustering of DM, what in turn results in a suppression of the matter power spectrum and, therefore, a lower value of $\sigma_8$. What is important about this is that the interaction does not affect the background and, therefore, the power suppression in the formation of structures is genuinely caused by sub-horizon physics and not at the expense of lowering the value of $\Omega_{\rm M}$. As a matter of fact, the preferred value of $\Omega_{\rm M}$ for the interacting model is slightly higher than that of $\Lambda$CDM. This clearly signals the importance of the interaction in order to lower $\sigma_8$ and how efficient the DM-DE coupling is for that. This result is then in the right direction to alleviate the $\sigma_8$ tension of Planck data and galaxy surveys. Our results are along the lines of the findings in \cite{PhysRevD.94.043518}.

The second remarkable result we obtain is that a vanishing coupling parameter is clearly excluded at more than 3$\sigma$. If taken seriously, this would imply a {\it detection} of an interaction between DM and DE and, as a consequence, the $\Lambda$CDM would be ruled out as compared to the interacting model. In other words, a Bayesian model comparison analysis would tend to favour the interacting model over $\Lambda$CDM. Notice that the interaction only introduces one additional parameter with respect to $\Lambda$CDM, so it should not be very penalised in the computation of the corresponding Jeffrey's factor. As a matter of fact, we find $\chi^2_{\Lambda{\rm CDM}}=11983.8$ and $\chi^2_{\rm interacting}=11960.7$ for the best fit values, what represents an improvement of $\Delta\chi^2\simeq23$. However, one should be cautious when claiming a detection of an interaction as the one obtained here, since one should include more observational tests before drawing a definitive conclusion as well as performing a more careful and exhaustive analysis. In any case, this result is very suggestive and motivates further exploration of these interacting models.

 It is also worth commenting on the preferred value of the optical depth that is higher for the interacting model than in $\Lambda$CDM (at more than 1$\sigma$) and, accordingly, the reionization redshift corresponds to an earlier epoch. Due to the degeneracy of the optical depth with the primordial scalar amplitude, the higher value for $\tau_{\rm reio}$ is accompanied by a larger $A_s$ so that the higher primordial power is compensated by a larger opacity during reionzation. This seems to be the preferred situation in the presence of the interaction with respect to $\Lambda$CDM. Since polarisation is more sensitive to the reionization epoch, including polarisation data could bring $\tau_{\rm reio}$ closer to the $\Lambda$CDM value.

%%%%%%%%%%%%%%%%%%%%%%%%%%%%%%%%%%%%%%%%%%%%%%%%%
%%%%%%%%%%%%%%%%%%%%%%%%%%%%%%%%%%%%%%%%%%%%%%%%%    
    \section{Discussion} \label{sec5}
   In this work we have explored an interacting model of DM and DE such that the background cosmology is not modified and the interaction only affects the perturbations when they enter the horizon. This is achieved by introducing a coupling between the dark components proportional to their relative velocity. Thus, assuming that they have a common large scale rest frame, only when peculiar velocities are developed the interaction can have an effect. Furthermore, the efficiency of the interaction, as in most interacting systems in cosmology, crucially depends on whether the interaction rate is larger or smaller than the Hubble constant. We have assumed a constant coupling that leads to a growing interaction rate with respect to the Hubble expansion rate so that the system inefficiently interacts in the early universe and only at late times the interaction becomes efficient. At this point, the DM-DE system behaves as a single fluid with a common velocity. This causes the growth of DM structures to cease and produces a suppression of the matter power spectrum on small scales which, in turn, allows for a lower value of $\sigma_8$ without modifying the background cosmological parameters and can alleviate the observed tensions. Interestingly, it is in fact possible to lower the value of $\sigma_8$ with respect to $\Lambda$CDM even for higher values of $\Omega_{\rm M}$ (keeping all other cosmological parameters fixed). Another interesting result that we have obtained is that the turnover of the matter power spectrum can be shifted towards larger scales for the same values of the cosmological parameters. This is a very distinctive signature of the interaction considered here different from most models where the turnover is robustly determined by the equality scale. Finally, since at recent times the DM is dragged by the DE component while baryons are not, this model also predicts the appearance of an additional motion on small scales between DM and baryons. This may affect the bias factor and, moreover, it may induce a mismatch between the center of the galaxies and that of the corresponding DM haloes.
   
   We have run a full MCMC to confront the predictions of the model to cosmological observations and we have obtained better agreement than $\Lambda$CDM. From our analysis, we have obtained nearly the same values for the cosmological parameters as in $\Lambda$CDM (within 1$\sigma$), which is consistent with the fact that we are not modifying the background cosmology. However, the improved fit arises because we have obtained a preferred lower value of $\sigma_8$, also in accordance with our expectations from the analytical analysis and the numerical results. Interestingly, we have obtained a preferred non-vanishing value for the interaction between DM and DE, with the vanishing coupling corresponding to $\Lambda$CDM excluded at more than $3\sigma$. Of course, it is not possible to make any strong claim at this point, but our results encourage to further investigate this family of interactions in the dark sector. As we have noticed several times throughout this work, some distinctive signatures of the interaction considered here are shared by models where the DM sector is coupled to a thermal bath of dark radiation independent of the DE sector. These models can further alleviate the $H_0$ tension due to the presence of extra-radiation at the background level (see e.g. \cite{Ko:2016uft,Ko:2017uyb,Buen-Abad:2017gxg}). Some of the similarities can be attributed to the fact that we have assumed DE to have $c_{s\mathrm{(DE)}}=1$ so it (nearly) behaves as a relativistic fluid at the perturbative level, but (nearly) as a cosmological constant at the background level. It would be interesting to study to what extent both frameworks could be merged or combined so that the tensions are alleviated without resorting to additional radiation components, but rather a modified DE sector. This is left for future work.

\acknowledgments
We thank the Research Council of Norway for their support. JBJ acknowledges support from the  {\textit{ Atracci\'on del Talento Cient\'ifico en Salamanca}} programme and the MINECO's projects FIS2014-52837-P and FIS2016-78859-P (AEI/FEDER). This paper is based upon work from the COST action CA15117 (CANTATA), supported by COST (European Cooperation in Science and Technology). The simulations were performed on resources provided by 
UNINETT Sigma2 -- the National Infrastructure for High Performance Computing and 
Data Storage in Norway.

\bibliographystyle{JHEP}

\end{document}